\documentclass[preprint]{aastex631}

\newcommand{\ISOIS}{IS$\odot$IS}
\usepackage{soul}
\usepackage{subfigure}
\usepackage{multirow}
\revised{Feb 12, 2026}

\submitjournal{ApJ}

\begin{document}

\title{Parker Solar Probe observations of solar energetic particle (SEP) events with inverse velocity arrival (IVA) features}

\correspondingauthor{Zigong Xu}
\email{zigongxu92@gmail.com, zgxu@caltech.edu}

\author[0000-0002-9246-996X]{Z. G. Xu}
\affiliation{California Institute of Technology, MC 290-17, Pasadena, CA 91125, USA}
\author[0000-0002-0978-8127]{C. M. S. Cohen}
\affiliation{California Institute of Technology, MC 290-17, Pasadena, CA 91125, USA}
\author[0000-0002-0156-2414]{R. A. Leske}
\affiliation{California Institute of Technology, MC 290-17, Pasadena, CA 91125, USA}
\author[0000-0003-0581-1278]{G. D. Muro}
\affiliation{California Institute of Technology, MC 290-17, Pasadena, CA 91125, USA}
\author[0000-0002-3840-7696]{A. C. Cummings}
\affiliation{California Institute of Technology, MC 290-17, Pasadena, CA 91125, USA}
\author[0000-0002-4559-2199]{O. M. Romeo}
\affiliation{Space Sciences Laboratory, University of California, Berkeley, CA 94720, USA}

\author[0000-0002-3176-8704]{D. Lario}
\affiliation{NASA Goddard Space Flight Center, Greenbelt, MD 20771, USA}
\author[0000-0001-6160-1158]{D. J. McComas}
\affiliation{Department of Astrophysical Sciences, Princeton University, Princeton, NJ 08544, USA}
\author[0000-0002-7341-2992]{M. E. Cuesta}
\affiliation{Department of Astrophysical Sciences, Princeton University, Princeton, NJ 08544, USA}
\author[0000-0003-2847-7110]{S. Pak}
\affiliation{Department of Astrophysical Sciences, Princeton University, Princeton, NJ 08544, USA}
\author[0000-0003-0412-1064]{L. Y. Khoo}
\affiliation{Department of Astrophysical Sciences, Princeton University, Princeton, NJ 08544, USA}
\author[0000-0001-7952-8032]{H. A. Farooki}
\affiliation{Department of Astrophysical Sciences, Princeton University, Princeton, NJ 08544, USA}
\author[0000-0002-3093-458X]{M. M. Shen}
\affiliation{Department of Astrophysical Sciences, Princeton University, Princeton, NJ 08544, USA}
\author[0000-0002-0972-8642]{S. Kasapis}
\affiliation{Department of Astrophysical Sciences, Princeton University, Princeton, NJ 08544, USA}
\author[0000-0003-2134-3937]{E. R. Christian}
\affiliation{NASA Goddard Space Flight Center, Greenbelt, MD 20771, USA}
\author[0000-0003-1960-2119]{D. G. Mitchell}
\affiliation{Johns Hopkins University Applied Physics Laboratory, Laurel, MD 20723, USA}
\author[0000-0002-4722-9166]{R. L. McNutt}
\affiliation{Johns Hopkins University Applied Physics Laboratory, Laurel, MD 20723, USA}
\author[0000-0001-6589-4509]{A. Kouloumvakos}
\affiliation{The Johns Hopkins University Applied Physics Laboratory, MD 20723, USA}
\author[0000-0003-4501-5452]{J. Grant Mitchell}
\affiliation{Johns Hopkins University Applied Physics Laboratory, Laurel, MD 20723, USA}
\author[0000-0001-6010-6374]{G. D. Berland}
\affiliation{Johns Hopkins University Applied Physics Laboratory, Laurel, MD 20723, USA}
\author[0000-0002-3737-9283]{N. A. Schwadron}
\affiliation{University of New Hampshire, Durham, NH 03824, USA}
\author[0000-0002-2825-3128]{M. E. Wiedenbeck}
\affiliation{Jet Propulsion Laboratory, California Institute of Technology, Pasadena}
\author[0000-0002-7728-0085]{M. L. Stevens}
\affiliation{Smithsonian Astrophysical Observatory, Cambridge, MA 02138, USA}
\author[0000-0003-2079-5683]{R. C. Allen}
\affiliation{Southwest Research Institute, San Antonio, TX 78238, USA}


\begin{abstract}
In SEP events, velocity dispersion (VD) is characterized by the earlier arrival of faster, higher-energy particles relative to slower ones, assuming negligible acceleration time and transport effects. This characteristic enables estimation of the time of particle release near the Sun. The "Labor Day event" at Parker Solar Probe (PSP) on 2022 September 5 provided a unique arrival profile, in which the
medium energy ($\sim$ few MeV) particles arrive earlier than both lower and higher energy particles. This created a so-called "nose" structure in the intensity spectrogram formed by measurements from the two energetic particle instruments, EPI-Lo and EPI-Hi, of the Integrated Science Investigation of the Sun (\ISOIS) suite. Unlike typical VD,
the delayed arrival of higher energy particles compared to medium energy particles, i.e., ”inverse velocity arrival” (IVA), could be caused by various acceleration, transport and instrumental effects, including shock acceleration. The shock requires time to accelerate particles to higher energies; the connectivity between the spacecraft and the particle source modify arrival time profiles; and even the instrument sensitivity modifies the time profile since higher energy particles are generally less abundant than lower energy particles.
By applying a new method based on the contour-line of the intensity, we found 14 IVA events in the \ISOIS ~ observations up to the end of 2024.
Several parameters that may modify velocity dispersion characteristics are further explored including the spacecraft radial distance, the speed of corresponding CMEs and shocks, the angle between the shock normal and the upstream magnetic field, and the spacecraft magnetic footpoint longitudinal separation from the flare location. The energy of the early arriving particles, i.e., the nose energy, can be grouped into low (L, $<$~0.5 MeV), medium(M, 0.5 - 5 MeV), and high(H, $>$~5 MeV) categories. Most (11/14) of the IVA events have medium nose energies. We emphasize the importance of instrumental effects on the shape of the IVA pattern.
This SEP list provides ingredients for examination of shock acceleration in the inner heliosphere, and the existence of IVA events sheds new light on the acceleration and propagation of SEPs.

\end{abstract}



\section{Introduction}

Solar energetic particles (SEPs) are primarily accelerated by flare eruption related magnetic reconnection processes or by shocks driven by coronal mass ejections (CMEs) via drift or diffusive shock acceleration \cite[and references therein]{reames_two_2013, reames1998SSRv, desai_large_2016}.
The CME-related energetic particles can spread widely in longitudinal extent and have been observed at multiple locations in the inner heliosphere. The energy of the particles ranges from tens of keV to hundreds of MeV, and even GeV.
SEP events' wide spread in space and broad energy range make them one of the most concerning hazards of space exploration, causing significant space weather events that interrupt human activities in the near-Earth space \citep{guo_radiation_2021, guo_particle_2024, cucinotta_cancer_2024}.

Although numerous studies, including both simulations and observations, have been conducted \cite[see review by ][and references therein]{desai_large_2016}, the acceleration and transport of particles in space are still poorly understood. For instance, most of our in-situ observations are conducted from a distant location at 1 au. The long distance between the particle source and the observer can introduce considerable transport effects, which can blur the initial particle signatures from the acceleration process(es). It is difficult to separate particle signatures of transport from those due to acceleration without observations of the accelerated particles closer to the source location and at an earlier phase in the process.

%


Thanks to its unique orbit around the sun, Parker Solar Probe (PSP, \cite{Fox_Solar_2016SSRv}) offers an opportunity to closely scrutinize the acceleration processes at CME-driven shocks. One special event is the well-studied Labor Day event on 2022 September 5, and its associated extremely fast CME \citep{cohen_observations_2024, romeo_near-sun_2023, paouris_space_2023, trotta_properties_2024}.  This event occurred when PSP was near perihelion at an unprecedented heliocentric distance of 0.07 au (15.45 solar radii).
The onset of this SEP event exhibits a completely different characteristic as compared to our more typical observations of SEP events.

\begin{figure}
    \centering  
    \subfigure[]{
    \label{Fig:1a}
    \includegraphics[width=0.8\linewidth]{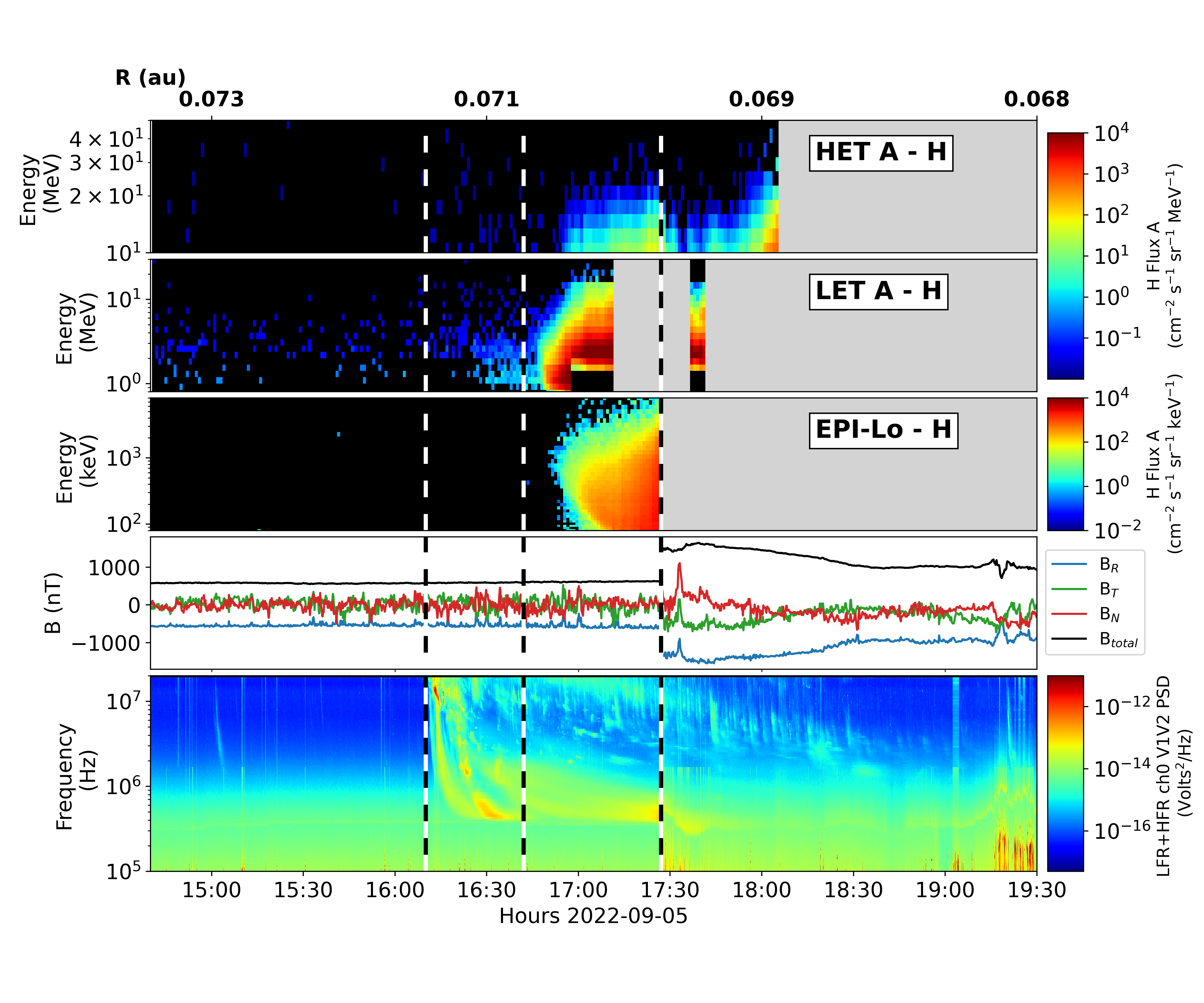}
    }
    \subfigure[]{
    \label{Fig:1b}
    \includegraphics[width=0.5\linewidth]{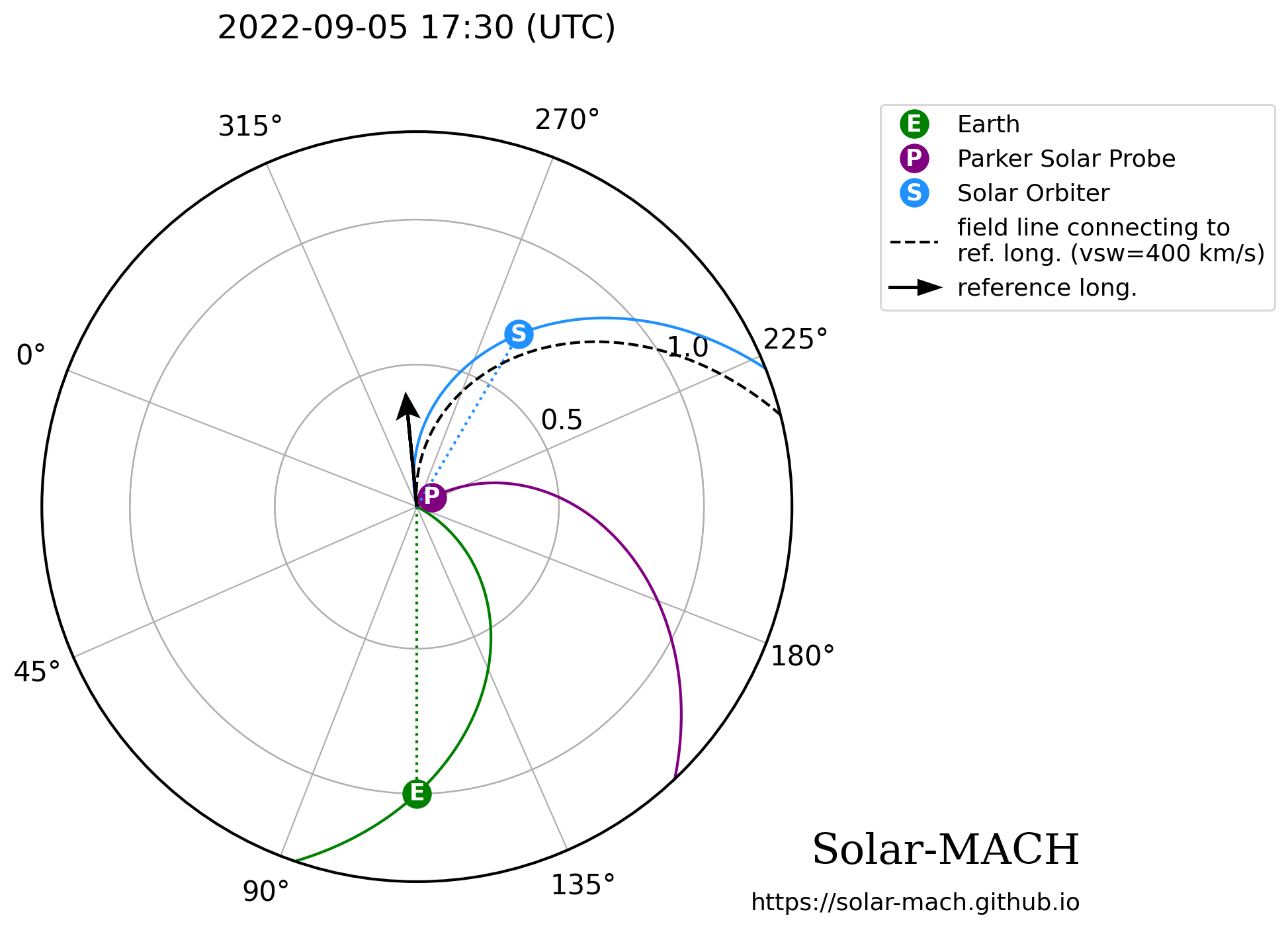}}
    \caption{(a): PSP observations of the energetic particles from EPI-Lo, EPI-Hi/LET + HET, the magnetic field variation and the radio emission by FIELDS \citep{bale_fields_2016} before and during the Labor Day event on 2022 September 5. Three vertical lines, from left to right, mark the flare eruption time, SEP event onset time \citep{cohen_observations_2024}, and the shock arrival time, respectively. (b): The relative positions of PSP and Solar Orbiter to the flare region during the event \citep{Gieseler2023FrASS}. The Parker spiral lines are calculated using the averaged solar wind velocities measured at PSP before the start of the SEP events.}
    \label{fig:2022-0905}
\end{figure}

Based on Figure 3 and 4 from \cite{cohen_observations_2024}, we reproduce PSP observations during the 2022 September 5 event in Figure~\ref{Fig:1a} as an overview of this event, where energetic particle measurements by the Integrated Science Investigation of the Sun (\ISOIS; \cite{mccomas_integrated_2016}) including both EPI-Lo \citep{hill_mushroom_2017} and EPI-Hi (i.e., LET and HET; \cite{wiedenbeck_capabilities_2017}) are displayed in the first three panels. The three components of the magnetic field vector \citep{bale_fields_2016} in RTN coordinates and the total magnetic field strength are presented in the fourth panel. The fifth panel of Figure~\ref{Fig:1a} show the radio emission measured by the radio wave antennas of the FIELDS instrument \citep{pulupa_solar_2017}. Three vertical dashed lines from left to right indicate the start of the flare eruption and the type-III radio burst at around 16:10 UT, the start of the SEP event at around 16:42 UT, which is based on the observation of EPI-Hi/LET, and the arrival of a shock at around 17:27 UT, respectively.
We show the complete measurements from both EPI-Lo and EPI-Hi, although they overlap in their energy coverage.

The striking feature of this event is the temporal onset profile of energetic particles across a broad energy range, where high energy particles (above a few MeV) and lower energy particles (below 500 keV) arrive later than the medium energy particles of $\sim$ few MeV energies leading to a nose shape in the dynamic energy spectrogram shortly before the shock arrival, as shown in the second and third panels of Figure~\ref{Fig:1a}. In other words, the normal velocity dispersion (VD) exists below 1 MeV but appears inverted above a few MeV in this SEP event.

This phenomenon challenges our understanding of the initial phase of an SEP event, which typically either has an obvious VD or has a blunt and blurred onset without a clear VD.  Dispersion in the onset time is expected under the assumption that the SEPs at different energies are released at the same time and travel roughly scatter-free along the same magnetic field line, resulting in the faster particles arriving earlier than the slower ones.
Many studies have commonly fit the VD, i.e., velocity dispersion analysis (VDA), and utilized these assumptions to determine the release time of the particles and then relate that to potential solar activity \citep[e.g.,][]{saiz_estimation_2005, Huttunen_Proton_2005AA, zhao_statistical_2019, laitinen_correcting_2015,Xu2020ApJ, Xu_2024ApJ, muro_radial_2025}. This analysis is most suitable for events with abrupt and highly anisotropic onsets \citep{wang_estimation_2015}, which suggests minimum scattering of the particles and presumably good magnetic connectivity between the observer and the source. Meanwhile, the fitted path length of the interplanetary magnetic field that particles travel along potentially reveals information regarding the particle transport. For instance, the path length will be larger when the SEP event is observed inside the flux rope structure of an interplanetary coronal mass ejection (ICME) \citep{wimmer-schweingruber_unusually_2023}.
Compared with the normal VD, this new inverse feature observed by PSP represents a more complicated scenario and can be caused by many mechanisms, as we will discuss. Since it is not simple diffusive transport of particles along the magnetic field lines (in the sense of VD), instead of calling it inverse velocity dispersion (IVD, \citep{ding_investigation_2025, kouloumvakos_shock_2025, li_delayed_2025}, we propose to call the feature inverse velocity arrival (IVA, \cite{IVA_xu_AGU_2024}), which better reflects its complex physical origin, and avoids misleading interpretations. More importantly, we differentiate IVA from earlier observations of IVD occurring in the vicinity of Earth’s magnetosheath and upstream of the bow shock, which has been observed by multiple instruments \citep{anderson_measurements_1981, anagnostopoulos_magnetospheric_1986, sarris_simultaneous_1987, lee_inverse_2016,lee_mms_2017}. These observations of IVD are of magnetospheric origin caused by different acceleration mechanisms and in comparison to the observations we display later, exhibits very different energy spectrum shape.



As first described by \citet{cohen_observations_2024}, this IVA feature can be explained by a balance between the time required to accelerate and release particles at the CME-driven shock and the travel time of those particles from the shock front to the spacecraft. Since PSP was sufficiently close to the approaching shock when particles were released, the later-released higher energy particles, which take a longer time to gain energy, could not overtake the medium energy particles that were released earlier, before they reached the spacecraft, leading to the inverted velocity arrival in the higher energy range.

Recently, the IVA feature of the 2022 September 5 event has been simulated and reproduced in multiple studies. \cite{do_time-dependent_2025} investigated the temporal variation of the acceleration and escape of charged particles at a traveling shock. By solving a time-dependent 1D transport equation analytically, the nose structure of EPI-Lo was successfully generated and can be explained as the lack of higher energy particles that the young shock is unable to produce.
\citet{kouloumvakos_shock_2025} constructed a data-driven SEP model from the corona to reproduce observations from both PSP and Solar Orbiter \citep{Mueller-2020-SolO} for the same event, where the latter observed an extremely intense SEP event without the inverted trend at the start of the SEP event. They found that PSP was first connected to the weak, east flank of the CME with a subcritical (lower Mach number) shock, which later became supercritical (higher Mach number). The ongoing particle acceleration at the flank of the shock wave causes the IVA and the nose structure. Meanwhile, cross-field diffusion transport processes seem to worsen the development of the nose observed at PSP.

Meanwhile, Solar Orbiter's onboard Energetic Particle Detector (EPD) suite of instruments \citep{RodriguezPacheco-2019-EPD} also recorded several additional SEP events having the IVA feature \citep{li_delayed_2025, allen_delayed_2026}. 
More interestingly, for one SEP event on 2022 June 7, with a long-lasting inverted trend, \citet{ding_investigation_2025} utilized a three-dimensional (3D) SEP model to reproduce the observations and proposed that the primary driving factor of the IVA onset for this event is the magnetic connectivity between the observer and the non-uniform shock driven by the CME that originated the event. The higher-energy particles that arrive later are from a more efficient acceleration site at the shock which Solar Orbiter connected to later as it propagated outward. They also state that cross-field diffusion transport could affect the characteristics of such SEP events, i.e., the energy of the first arriving particles and the duration of such events.

It is worth noting that although both Solar Orbiter and PSP are currently operating in the inner heliosphere, their orbits are quite different. PSP gets much closer (with a closest distance of 0.04 au, 9.86 solar radii) to the Sun and can observe the initial acceleration stages of the shock during an SEP event, while SolO is mainly positioned between 0.3 and 1 au, with an increasing latitude recently out of the ecliptic plane. It is perhaps unsurprising that the previous studies from PSP and SolO yielded differing conclusions, given their distinct observational perspectives and instrument performance.
Moreover, the two instruments of \ISOIS ~onboard PSP have shown different particle characteristics at the start of an event. For example, in the Labor Day event, the onset of the inverse feature at EPI-Lo is later than that measured by EPI-Hi, and the slopes of the IVA part are also different, as shown in the panel (a) of Figure~\ref{fig:2022-0905}. The physical explanation behind this discrepancy might be due their different measurement sensitivity, as we discuss later.

Though the balance between particle acceleration and transport time scales offers a compelling explanation for the IVA, the above two issues also remind us that our understanding of the overall picture of such events is far from perfect, and the real cause of the inverse trend warrants further investigation.
Therefore, in this paper, we present a survey of energetic particle observations from PSP between 2019 and 2025, searching for more SEP events with IVA onsets. 

To consistently identify IVA events, for the first time we propose a quantitative method that uses contour lines to visualize and determine the inverse features. Following this method, a dataset that consists of 14 IVA events is generated. Despite the limited size of the list, we characterize the nose energies of IVA events, their relationship to the CME/shock and solar source region, and the corresponding CME and shock properties. Interestingly, based on the shape of the contour lines which show distinct patterns in the dynamic spectrogram onsets, we find that all SEP events can be divided into three groups, which are VD event, Nose-only event, and Mixed event. These features are described in detail in the observation section. The paper ends with a detailed discussion, and a brief conclusion summarizes our findings.





\section{Instrumentation}\label{sec:instru}

In this study, we focus on the SEP observations from PSP/\ISOIS ~ \citep{mccomas_integrated_2016}. We utilize proton measurements from both EPI-Lo and EPI-Hi, with the latter consisting of both LET and HET. Together, they cover the proton energy range from a few tens of keV to a few tens of MeV. EPI-Hi and EPI-Lo have an energy overlap between 1 MeV and a few MeV, where the data can be used to inter-calibrate the two instruments. Previous observations \citep{joyce_energetic_2020,cohen_pspisis_2021, mitchell_energetic_2021, cohen_observations_2024, Xu_2024ApJ} have already shown the general consistency of the two instruments during SEP events.
More instrumental details and recent updates of EPI-Hi's capabilities can be found in  \cite{mccomas_integrated_2016,  wiedenbeck_capabilities_2017, pak_species-dependent_2025}

Relevant to our study, we emphasize that EPI-Lo and EPI-Hi utilize distinct measurement techniques and have different sensitivity \citep{mccomas_integrated_2016}.
A single aperture of EPI-Lo has a small geometry factor of $\sim$7.7 $\times$ 10$^{-4}$ $cm^2\ sr$, and the whole instrument, which consists of 80 apertures, has a total geometry factor of $\sim$ 0.061 $cm^2\ sr$, for particles with energies below $\sim$1 MeV. While EPI-Hi, including both LET and HET, measure particles of energies above MeV/nuc with an overall geometry factor around $\sim$ 0.5 $cm^2\ sr$.  The precise values vary with energy and for different operation modes, i.e., dynamic threshold (DT) modes which are employed when high intensities could cause a significantly reduced livetime \citep{cohen_pspisis_2021, cohen_observations_2024}. 
Despite the different geometry factors of the instruments, an IVA event can appear over a broad energy range spanning both EPI-Lo and EPI-Hi. In some cases, the energy of the first-arriving particles can exceed 10 MeV; hence, we use data from EPI-Lo along with LET and HET for this study.

\section{Observations}


\begin{figure}
    \centering
    \subfigure[]{
    \label{fig:contourline-spectrogram-a}
    \includegraphics[width =0.7\textwidth]{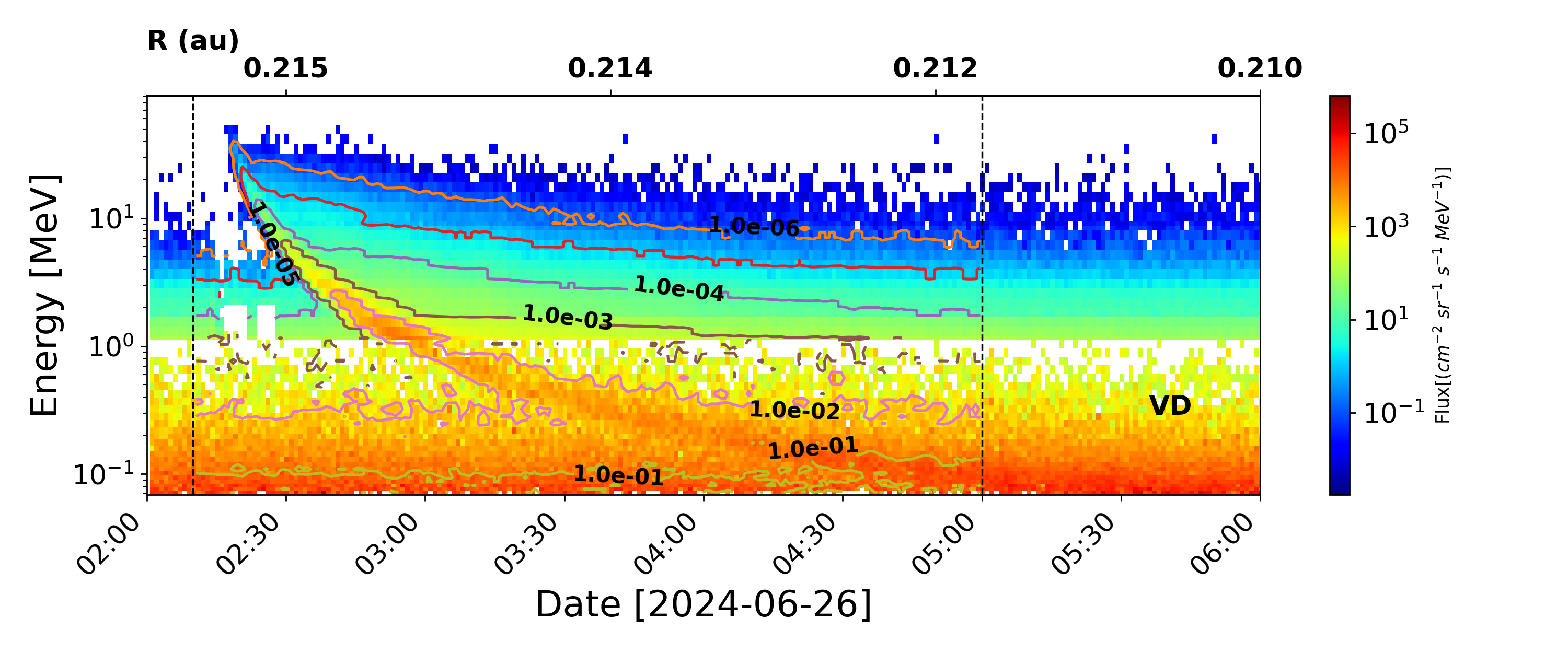}}
    \subfigure[]{
    \label{fig:contourline-spectrogram-b}
    \includegraphics[width = 0.7\textwidth]{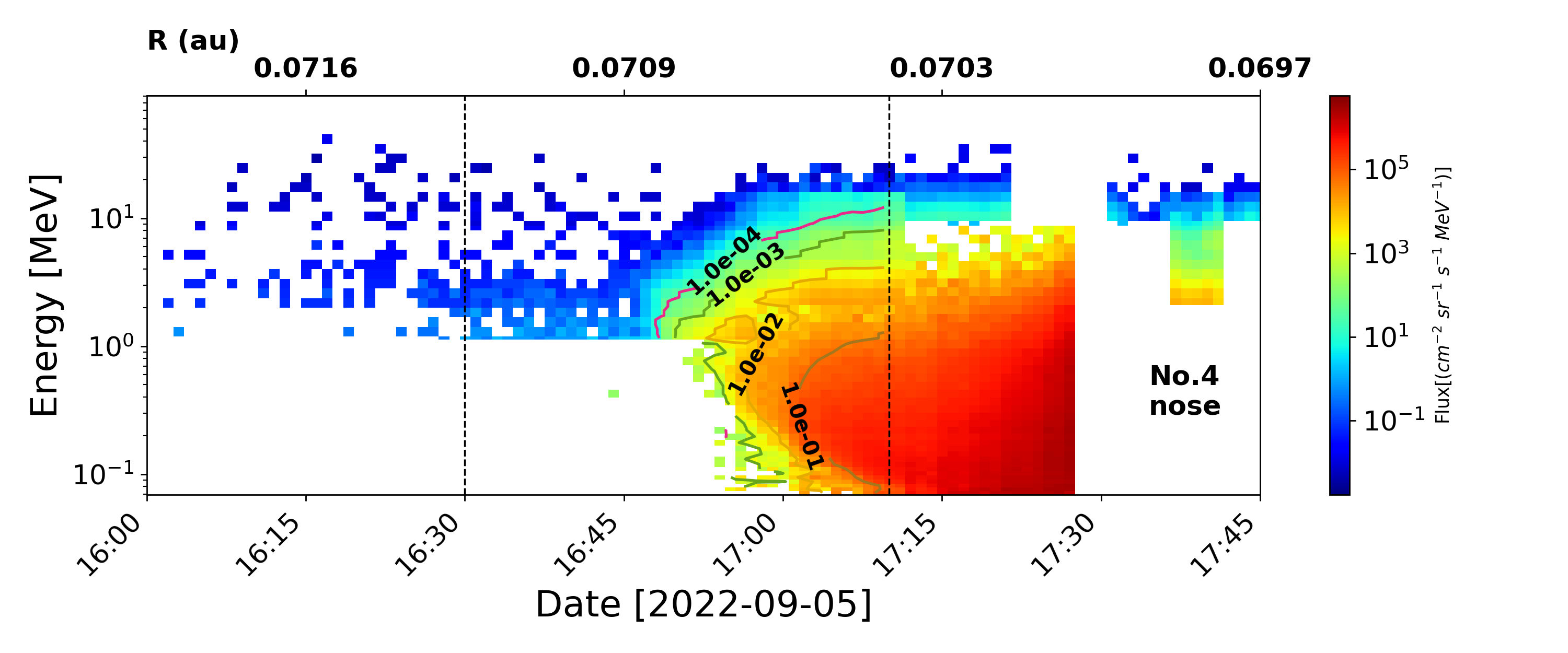}}
    \subfigure[]{
    \label{fig:contourline-spectrogram-c}
    \includegraphics[width = 0.7\textwidth]{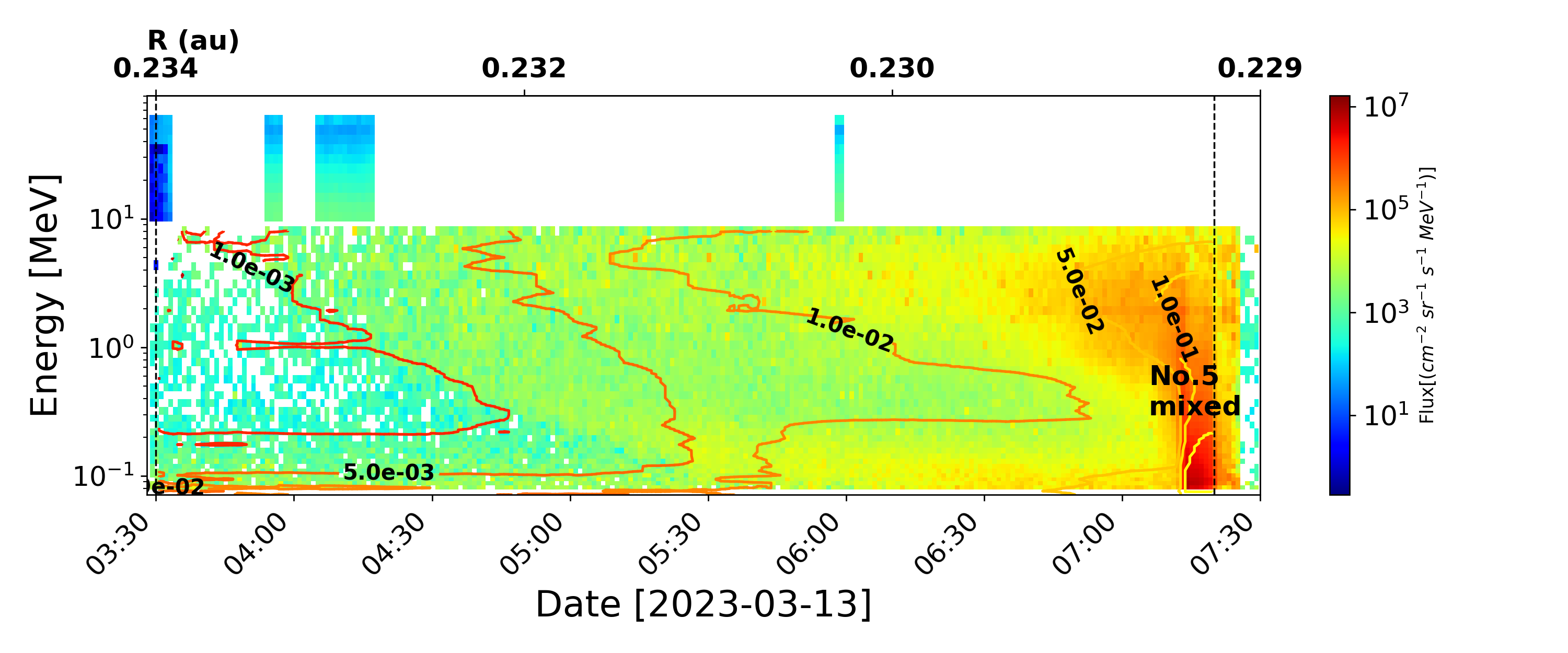}}
    
   \caption{Three types of SEP events (top to bottom): (a): VD event - Spiky event with normal velocity dispersion; (b): Nose-only event - SEP event with IVA features; (c): Mixed event - a mixture of the above two types with clear separation between the first arriving VD and the later inverse features. The intensity unit is $\mathrm{count}/(\mathrm{s\cdot sr\cdot MeV \cdot cm^2)}$. These plots combine measurements from EPI-Hi/LET-A and HET-A, and the mean intensity averaged over all EPI-Lo apertures. The contour lines are calculated between the two vertical dashed lines.}
    \label{fig:contourline-spectrogram}
\end{figure}

\subsection{The contour line method}

The top three panels in Figure~\ref{Fig:1a} show the energetic particle observations separately from EPI-Lo, LET, and HET, using individual color bars and units for ease of examining each data set independently. But it is difficult to directly compare their observations from such a figure, especially for features spanning the partially overlapping energy channels of EPI-Lo and LET. Therefore, in the rest of the manuscript, we display the observations from both EPI-Lo and EPI-Hi in the same format with the same intensity units, as given in the combined spectrograms in Figure~\ref{fig:contourline-spectrogram}. Because of the better counting statistics as compared to EPI-Lo, LET data are used within the overlapping energy channels of EPI-Lo and LET in the combined spectrogram, i.e., for energy $>$ 1 MeV, which varies with the change of the DT mode of LET. Below that energy, EPI-Lo data fill the spectrogram. The averaged intensities of LET and HET are employed within the shared energy channels of LET and HET, since they have similar fields of view and generally consistent intensities.


A big challenge when searching for more IVA related SEP events is how to consistently define and identify them. This is caused by several issues. 
First, the nose shapes vary across different instruments, particularly when plotted in the standard separate spectrogram plots.
As Figure~\ref{fig:2022-0905} shows, in the Labor Day event,  EPI-Lo and EPI-Hi seem to have different nose features, which manifest as different onset times in the similar energy channels, different nose energy, and even different slopes of the onset edge. With its larger geometry factor and thus higher sensitivity, EPI-Hi data are preferred for identifying the nose features. However, some low-energy events are only observed by EPI-Lo.
Additionally, when visually inspecting the standard spectrograms for IVA events, we found that many events display inverse arrival features, but they are preceded by a separate particle population exhibiting traditional VD. These 'mixed' events consist of two populations, with the nose and IVA marking the arrival of a second population, making it distinct from 'nose-only' events such as the 2022 September 5 event.
This type of event is more evident in the EPI-Hi observations as often the first particles population has significantly lower intensities than the nose population.

To overcome the limitations of the visual inspection method, we propose a "contour line method", as shown in Figure~\ref{fig:contourline-spectrogram}, to consistently identify the IVA events and determine the properties of the nose features. In the combined instrument spectrogram, we draw contour lines along constant particle intensity values for different fractional levels of 
This, along with the combined spectrogram with common intensity units, mitigates issues of relative instrumental sensitivity, which is apparent when the particle intensities are very low or below the instrumental 1-count threshold.  Increased accumulation time can also mitigate this effect, but this is not always an option when examining a transient effect, such as in this study. This concept is somewhat similar to the fraction used in the fractional velocity dispersion analysis method \citep{zhao_statistical_2019}, despite the fact that their fraction is defined relative to the peak intensity of each separate energy channel.  We choose a much lower intensity level that is similar to \cite{saiz_estimation_2005}. Contour lines are calculated for the combined spectrogram across different instruments, rather than being restricted to a single measurement.

In most of the cases that we analyze, we utilize the contour line of 0.1$\%$ (10$^{-3}$) of the peak intensity as a standard line for IVA determination. In some cases with high peak intensities and excellent statistics, we can use a lower flux line like 0.01$\%$ (10$^{-4}$).  For each event, the peak intensity is determined across the full combined EPI-Lo+EPI-Hi energy range and the selected time period.  Each time period spans from the start of the first arriving particles of the nose population to just before the shock arrival, or covers the peak of the intensity if no shock arrives.

As shown in panel (b) of Figure~\ref{fig:contourline-spectrogram}, our contour line method demonstrates reliable performance for the Labor Day event, and successfully captures the overall trend of the inverted edge. The two dashed lines indicate the time period for which the contour lines were calculated. We show multiple contour lines, with the factors listed along the lines, from 10$^{-4}$ to 10$^{-1}$ of the peak intensity. 
The nose pattern is clearly evident in multiple contour lines across both EPI-Lo and EPI-Hi energies although the characteristics of the nose feature may be slightly different depending on which contour line is selected.
While the spectrogram showed here is the particle intensity, the contour line method can also be applied to an energy flux spectrogram in which the particle flux is multiplied by the square of the energy with similar, consistent results regarding the identification of the IVA events.

For this study, we limited our selections of IVA events to those that were SEP events, rather than particle intensity increases associated with a corotating interaction regions (CIRs) or stream interaction regions (SIRs) by checking the corresponding observation of solar wind and magnetic field, as well as solar activities. Our survey found that some observed CIR/SIR proton intensity enhancements may also exhibit an inverse trend 
\citep{allen_radial_2021}; however, generally the shapes are different from those found in SEP events, i.e., there is no lower energy VD portion (critical to creating the nose feature) possibly because of their different acceleration and transport processes. While interesting, discussions of the IVA in CIR/SIR events are out of the scope of this paper. As part of our selection criteria, we require that the contour lines have a clear normal VD below the nose energy, although in some cases, the full VD might be missed due to observation limitations of EPI-Lo.
Based on our survey, we found that SEP events can be categorized into three groups with different contour patterns. We give three examples in Figure~\ref{fig:contourline-spectrogram}, and the dynamic spectrogram together with the contour lines for the rest of the SEP events with IVA features are provided in Figure~\ref{fig:Appendix-IVA}. We describe the three types and their characteristics in the following section.




\subsection{Three typical cases} \label{sec:threetypical}
Three SEP events are taken as examples of the three categories of SEP events, as shown in Figure~\ref{fig:contourline-spectrogram}. These are the VD event on 2024 June 26 (Figure~\ref{fig:contourline-spectrogram-a}), the nose-only SEP event on 2022 September 5(Figure~\ref{fig:contourline-spectrogram-b}), and the mixed event on 2023 March 13 (Figure~\ref{fig:contourline-spectrogram-c}). It should be noted, however, that the categorization can be somewhat subjective, particularly for nose-only and mixed event types. 

\subsubsection{VD event: 2024 June 26}
VD events represent a large group of SEP events that only show normal velocity dispersion. An example we give here is an ideal case which occurred on 2024 June 26, as shown in the top panel of Figure~\ref{fig:contourline-spectrogram}. This  "spiky event" has a sharp and abrupt onset in the particle intensities across a broad energy range. The intensity peak occurs soon after the event starts and is followed by a gradual decrease in the intensity back to pre-event levels.  The 5 contour lines are overlaid in the plot and show clear velocity dispersion without any inverse features. 
Notably, another characteristic of this event is its shorter duration at different energy levels than that of other typical events. This is why we also called it a spiky event, as it appears as a sharp spike in the temporal intensity profile, although this is not a common property of the VD events. 

\subsubsection{Nose-only event: 2022 September 5} \label{sec:nose-only}
The Labor Day event is what we identify from the contour lines as a nose-only event, as the dynamic spectrogram only displays a nose feature at the onset of the event. The inverted trend appears in most of the contour lines depicted in the middle panel of Figure~\ref{fig:contourline-spectrogram}. The VD is clear in the EPI-Lo measurements at the level of $10^{-3}$ of peak intensity, which is the standard contour line we use.
At the lower contour levels, below the sensitivity of EPI-Lo, the IVA is clearly shown in EPI-Hi/LET, but the normal VD is missed since LET lacks reliable measurements below 1 MeV.  The combined EPI-Lo and LET measurements allow the nose feature to be determined, and we estimate the nose energy via the $10^{-3}$ contour line.  It should be noted, however, that it is challenging to provide a precise nose energy, due to its broad energy spread.
Due to the enhancement of the particle intensity, the DT mode is triggered for EPI-Hi/LET before the time of the shock arrival at PSP at 17:27 UT, resulting in an empty region at the energy of a few MeV after the second dashed line in Figure~\ref{fig:contourline-spectrogram-b}. For more details of this event, see \cite{cohen_observations_2024}.

\subsubsection{Mixed event: 2023 March 13}

The last type of SEP event is the mixed event, which is the combination of two particle populations, one of each of the other two types. The first population exhibits normal VD and is followed by a second population whose onset appears as a nose feature. The example we provide here is the 2023 March 13 SEP event, whose spectrogram is shown in the bottom panel of Figure~\ref{fig:contourline-spectrogram}.
The normal VD is manifested as the 10$^{-3}$ contour line, and the nose feature is marked by the 10$^{-2}$ contour line. Unfortunately, the high particle intensity caused EPI-Hi to be in DT mode most of the time, resulting in the data gap shown in the proton spectrogram. Therefore, we do not have the detailed shape of the profile for the higher energy particles in the spectrogram \citep[see also][]{dresing_reason_2025}.
But the contour line of 10$^{-2}$ of the peak intensity still indicates a turnover at around a few MeV as well as the inverse trend in the energy region above it; therefore, it is reasonable to assume this line would continue to the higher energy region up to tens of MeV. 

Overall, PSP observations for this event indicated an intense SEP event with a sharp onset and an intensity peak associated with the passage of the CME-driven shock.
As reported by \cite{dresing_reason_2025}, PSP was close to the Sun and well connected to the eruption. The source of this intense SEP event was identified as a combined shock driven by multiple CMEs propagating in different directions.
Upon arrival at PSP, 
the shock was accompanied by an intense Energetic Storm Particle (ESP) event observed at proton energies $>$ 30 MeV, and was estimated to be very fast ($\sim$ 2800 $\mathrm{km/s}$) and parallel ($\theta_{Bn}\sim 8^{\circ}$), leading to one of the strongest shocks that have been observed in the heliospheric environment \citep{jebaraj_acceleration_2024}. 
\cite{jebaraj_acceleration_2024} also found that the electrons continue to be accelerated to higher energies while the shock propagated outward, which is highly unusual in SEP events.
A plausible scenario for this event is that the first population with normal velocity dispersion, which extends from tens of keV to tens of MeV, is from a well-connected CME-driven shock at the start of the event, whereas the second population with the IVA feature is due to the propagating and expanding shock that needs more time to accelerate particles to higher energies \citep{giacalone_source_2026}, as well as the energy-dependent escape of particles from the approaching shock \citep{lario_evolution_2019}.
Note that during this event, IVA features were also observed by BepiColombo \citep{benkhoff_bepicolombo_2021} and Solar Orbiter at the onset of the SEP event \citep{dresing_reason_2025, allen_delayed_2026}.



\subsection{IVA event list}
Using the contour line method and the criteria for the eligible SEP events, we obtained a list of SEP events with IVA features.
In total, we found 14 events, including the Labor Day event, in the period from August 2018 to the end of 2024 encompassing the first 22 orbits of PSP. These events are listed in Table ~\ref{tab:iva-list}.

The time in the table (second column) is defined as the moment when the nose appears at the selected contour line, which in most cases is the level of 10$^{-3}$ of the peak intensity. For the 2023 March 13 SEP event and the 2024 September 29 SEP event, we utilize different contour lines, $10^{-2}$ and $10^{-4}$ of the peak intensity respectively, to determine the nose time.
For a case like the Labor Day event, the nose times at different contour levels are similar. But for some cases, the start time of the nose can vary significantly for different intensity levels. Thus, the contour line that represents 10$^{-3}$ of the peak intensity within the selected region provides only a rough estimate of the time of appearance of the nose.

The other parameters in Table~\ref{tab:iva-list} include the distance of the spacecraft to the Sun when it recorded the event, the local speed of the CME-driven shock (V$_{sh}$) if the shock passes PSP, the sky-plane speed of CME (V$_{cme}$), the angle between the shock normal and the upstream magnetic field ($\theta_{Bn}$), the longitudinal separation between the magnetic footpoint of PSP and the flare location (D$_{lon}$) assuming a nominal Parker spiral and the measured solar wind velocity before the SEP event, the time the shock passes PSP given as a delay time in hours after nose time, the nose energy, and the instrument that observed the nose.
Given the proximity of PSP to the Sun for most of the events in the list, a Parker spiral is rather a good approximation for the spacecraft footpoint.

The shocks identified are the first shock that passed PSP after the start of the SEP event and arrived within 48 hours. Depending on the speed of the shock and the location of PSP, the time required for the shock to travel may vary.
Not all of the events have an accompanying shock passing PSP. As shown in the table, events 7, 8, and 9 did not exhibit any shock at PSP.
Additionally, there are two events (2 and 13)  for which the presence of a shock cannot be established owing to data gaps.
In the end, only 9 events have been identified as having associated shocks. The shock parameters of the Labor Day event and the event on 2023 March 13, are taken from the literature (as indicated in the table notes). 
The shock parameters, i.e., the $\theta_{Bn}$ and V$_{sh}$ of the remaining shocks, are determined using the Mixed-mode 3 method \citep{abraham-shrauner_interplanetary_1976, palmerio_mesoscale_2024}, which utilizes the magnetic field measured from PSP/FIELDS \citep{bale_fields_2016}, the solar wind density, and the solar wind speed. For each shock, we use the available solar wind data from PSP/SWEAP \citep{kasper_solar_2016} with a combination of the SPAN-I \citep{livi_solar_2022} velocity, SPAN-E \citep{whittlesey_solar_2020} density, SPC \citep{case_solar_2020} measurements, and electron density from the quasi-thermal noise (QTN) method. The V$_{\mathrm{CME}}$ are obtained from the CCMC DONKI\footnote{https://kauai.ccmc.gsfc.nasa.gov/DONKI/} dataset, and the results from literature \citep{khoo_multispacecraft_2024, paouris_space_2023, trotta_properties_2024, kouloumvakos_shock_2025, jebaraj_acceleration_2024}. 
 
The energy location of the nose, i.e., nose energy, may be related to different factors, for example, the distance of the shock from the spacecraft when the particles are released, the strength and the geometry of the shock \citep{do_time-dependent_2025, ding_investigation_2025}.
Note that determining the nose energy for each event is challenging since the nose features can be ambiguous and the nose energy can span an extended region of energy, especially during mixed events that have two populations, one with normal VD and the other with IVA. Here, we simply categorize the nose energy into three groups as a qualitative assessment: low ($<$ 0.5 MeV), medium (0.5 $<$ E$_{nose} <$ 5 MeV), and high ($>$ 5 MeV). In this list, most of the events (11) have medium nose energy, including the Labor Day event, where the nose energy is determined by  EPI-Hi rather than  EPI-Lo due to better statistics. Two events have high nose energies, and only one event on 2024 September 29, has low nose energy. The majority of the measurements of this event are from EPI-Lo below 1 MeV.



\setlength{\tabcolsep}{1pt}
\begin{table}[]
\small
    \centering
    \caption{A list of SEP events with the inverse velocity arrival features. 
    }
    \begin{tabular}{|c|c|c|c|c|c|c|c|c|c|}
    
    \hline
        No.           & Time           & Distance(au)  & $\theta_{Bn}(^o)$      & V$_{sh}$/V$_{cme}$(km/s) & D$_{lon}$\footnote{The longitudinal separation between the magnetic footpoint of SC, assuming solar wind speed of 400 km/s, and flare location. D$_{lon}$ Positive values mean that the footpoint is located to the west of the flare, and negative values indicate the east side.} & E$_{nose}$\footnote{Low(L): $\leq $ 0.5 MeV, Medium(M): (0.5, 5) MeV, High(H): $\geq$ 5 MeV} & dt$_{sh}(h)$  & Inst & Type \\
        \hline
        1   &2022-02-15 22:50&   0.38        &   30.62   &  1177.92 / 2315 \footnote{\cite{khoo_multispacecraft_2024}}        &        3.8&     M    &  8.58& Hi/Lo & mixed\\
        \hline
        2   &2022-08-27 03:30 &   0.39       &     --\footnote{'-' in 2 and 13 means the data gap in solar wind or mag data.}         & --      / 1372            &   -3.7&    M    &  --  & Hi & mixed\\
        \hline
        3   &2022-08-28 22:15&   0.34        &   72      & 774   /1148 &  -40.6&   M   & 4.70& Hi/Lo & mixed\\
        \hline
        \multirow{2}{*}{4}   & \multirow{2}{*}{2022-09-05 16:42}   &   \multirow{2}{*}{\textbf{0.07}}       &    \multirow{2}{*}{53 $^e$}   &       1360 \footnote{\cite{paouris_space_2023}}   , 1520 \footnote{\cite{trotta_properties_2024}}      &  \multirow{2}{*}{-62.1 }&   \multirow{2}{*}{M}     & \multirow{2}{*}{0.68}& \multirow{2}{*}{Hi/Lo} & \multirow{2}{*}{nose} \\
           &   &        &      &  /2480\footnote{\cite{kouloumvakos_shock_2025}}, 2039$^e$    &  &     & & &\\
        
        \hline 
        5$^*$   &2023-03-13 05:40&   0.23        &   8   &       2800 \footnote{\cite{jebaraj_acceleration_2024}}  /2127            &     -31.9                        &    M    & 1.58& Lo & mixed\\   
        \hline
        6   &2023-07-10 05:15&   0.54        &    88.20   & 473.26    /901      &       23.2                       &    M     &  38.75& Hi & mixed\\
        \hline
        7   &2023-07-16 07:30&   0.63       &      x\footnote{'x' symbol in 7, 8, and 9 means no shock arrived at PSP afterward. The nose of 7 is within a CME. }            &  x       /1441             &       -56.8                       &   M   &   x& Hi & mixed\\
        \hline
        8   &2023-11-09 15:30&   0.74       &     x            & x       /782            &       -39.7                      &   M    &   x& Hi & nose\\
        \hline
            9   &2024-01-01 01:00&   0.18       &    x               & x         /2184          &        -99.5                     &   H     &   x& Hi &mixed\\
        \hline
        10  &2024-05-10 21:30&   0.74       &   48.05       & 713.08  /1263         &         88.8                    &    M  &    40.63& Hi & nose\\
        \hline
        11  &2024-06-24 06:00&   0.27      &     57.21    & 462.11  /640        &        25.9                       &    M   &   16.33& Hi & nose\\
        \hline
        12$^{**}$  &2024-09-29 07:00&   \textbf{0.07}     &    56.89        & 594.53 /1148        &      118.8                      &     L  &     0.83& Lo & nose\\
        \hline
        13  &2024-10-09 03:10&   0.37    &       --            &      --      /1509        &      -64.5                    &    M &    --& Hi & mixed\\
        \hline
        14  &2024-10-24 04:35&   0.63     &     17.39       &   958.37  /1606    &         27.1                     &    H    &    18.75& Hi  &mixed \\
        \hline
        
    \end{tabular}\\
    \vspace{0.2cm}
    $^*$ used contour line 10$^{-2}$ of the peak intensity; $^{**}$ used contour line 10$^{-4}$ of the peak intensity; the rest used contour line 10$^{-3}$ of the peak intensity
    \label{tab:iva-list}
\end{table}

\begin{figure}
    \centering
    \includegraphics[width=0.4\textwidth]{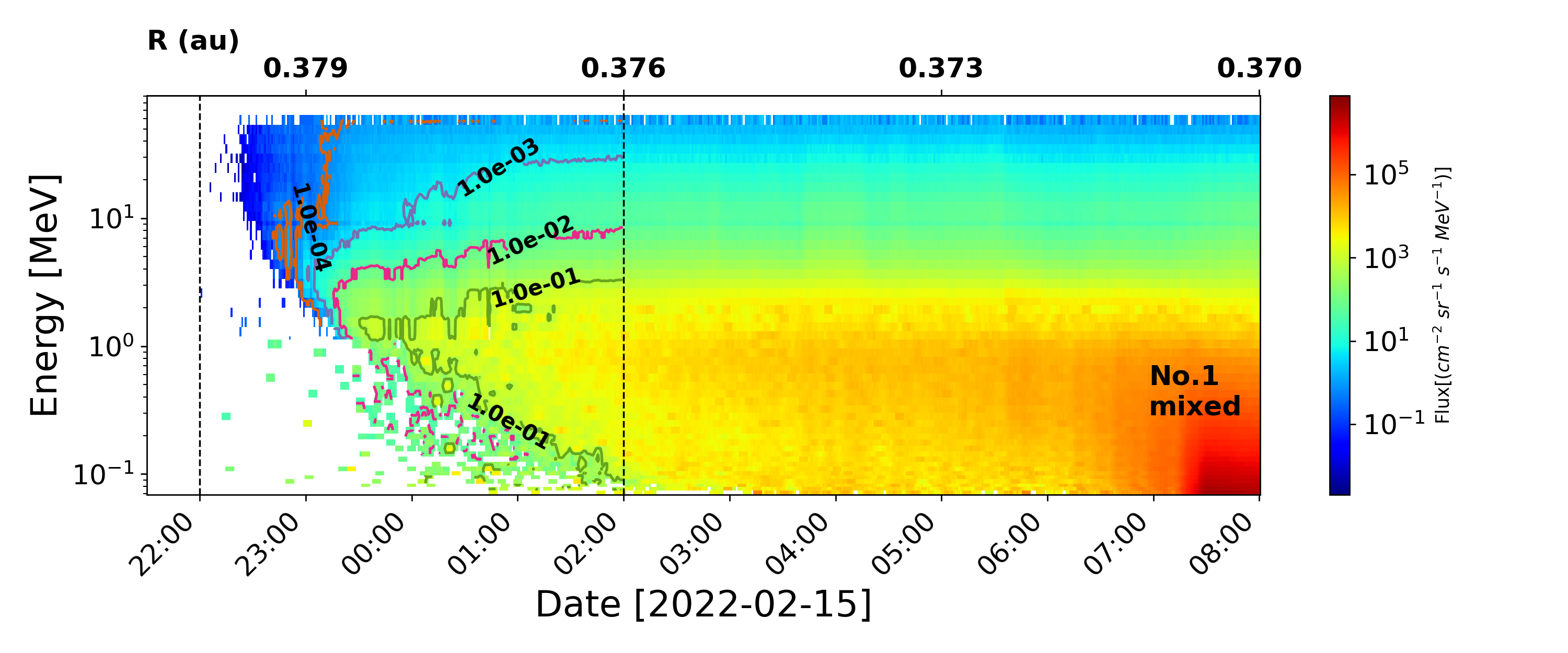}
    \includegraphics[width=0.4\textwidth]{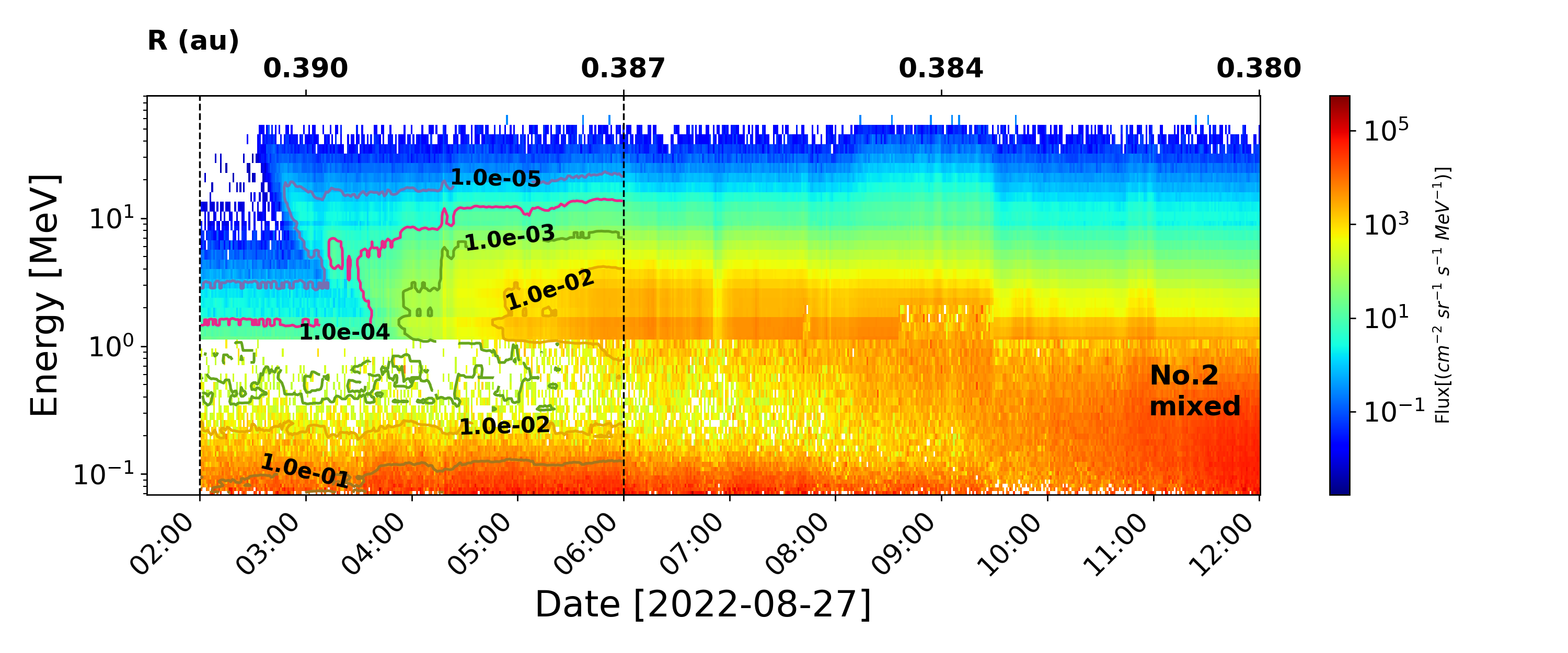}
    \includegraphics[width=0.4\textwidth]{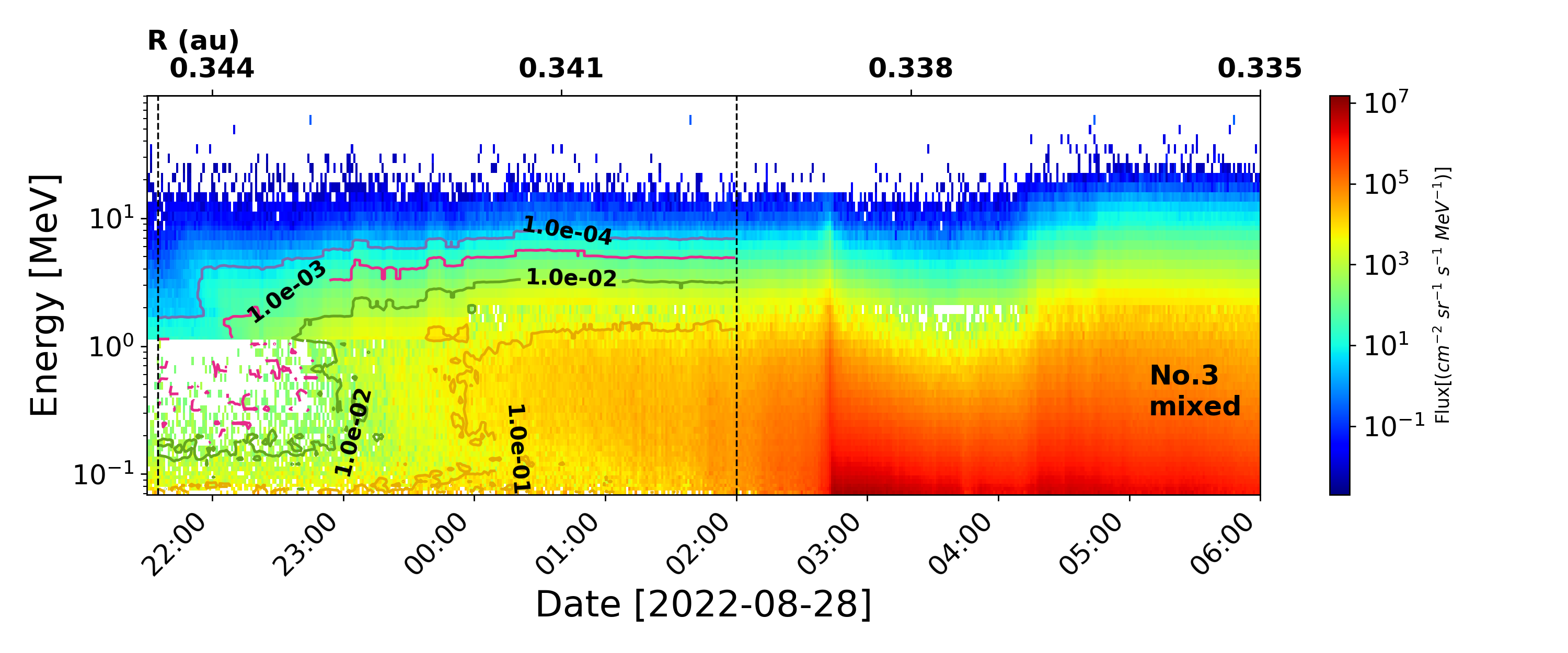}
    \includegraphics[width=0.4\textwidth]{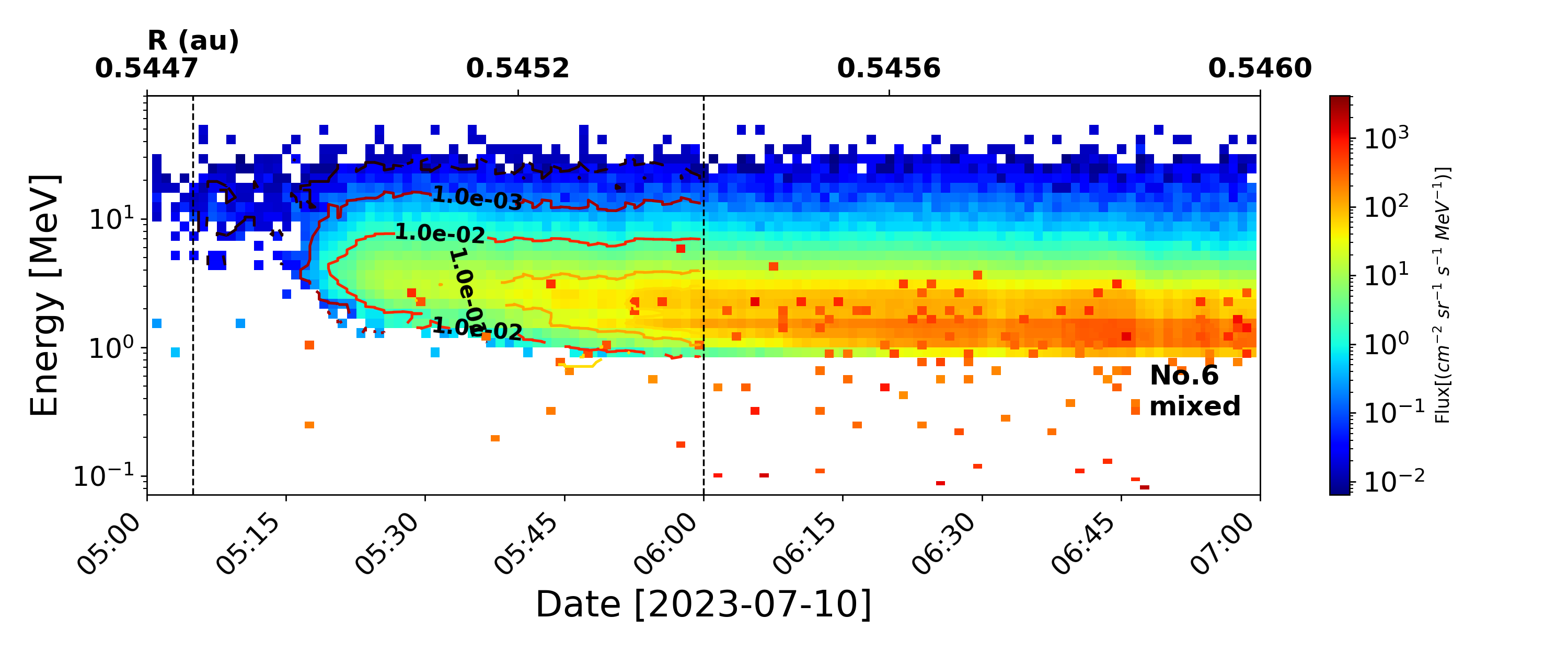}
    \includegraphics[width=0.4\textwidth]{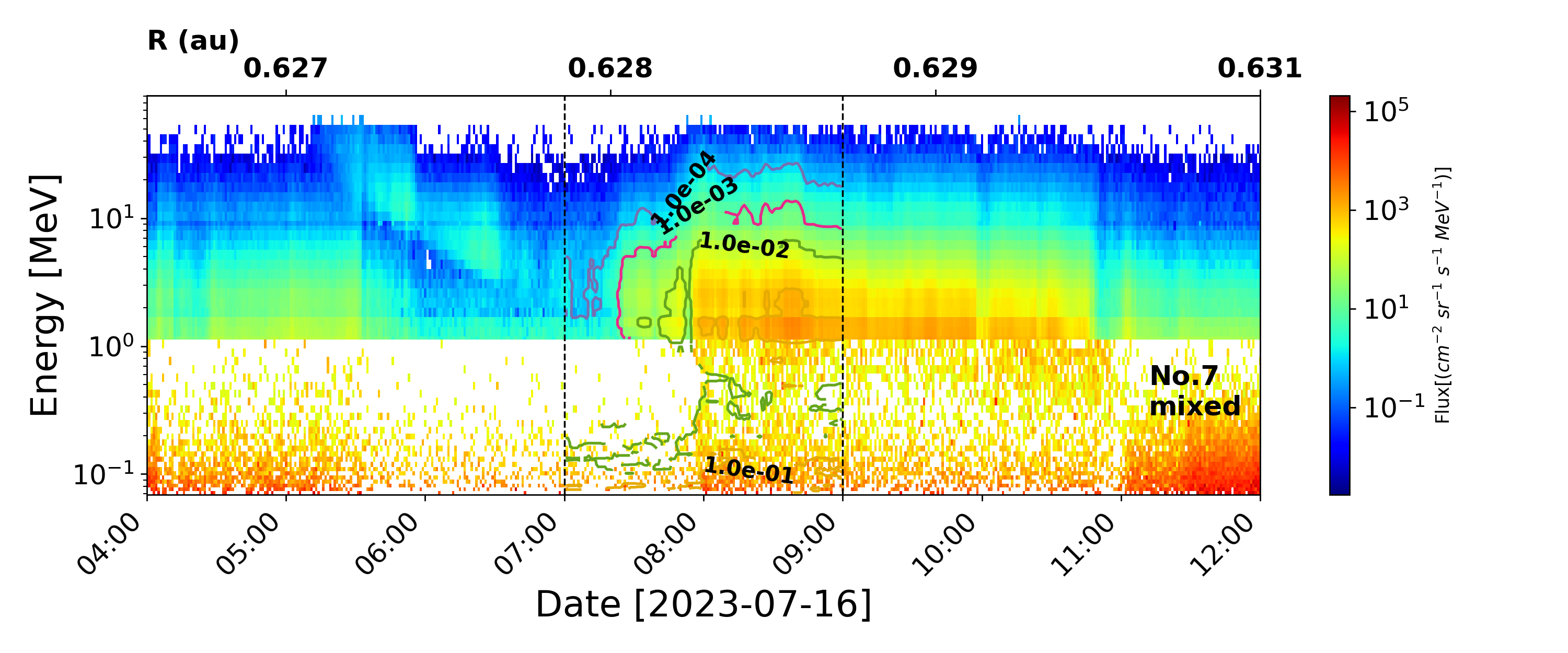}
    \includegraphics[width=0.4\textwidth]{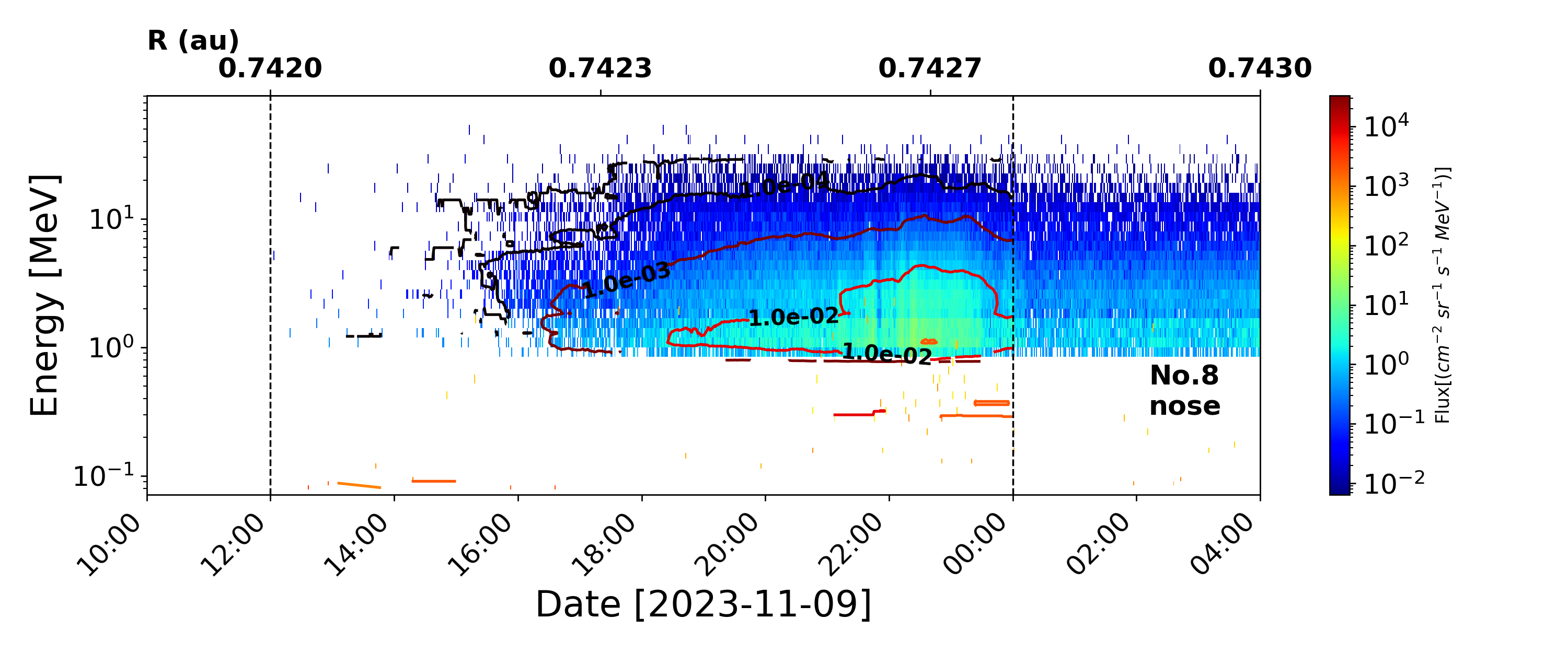}
    \includegraphics[width=0.4\textwidth]{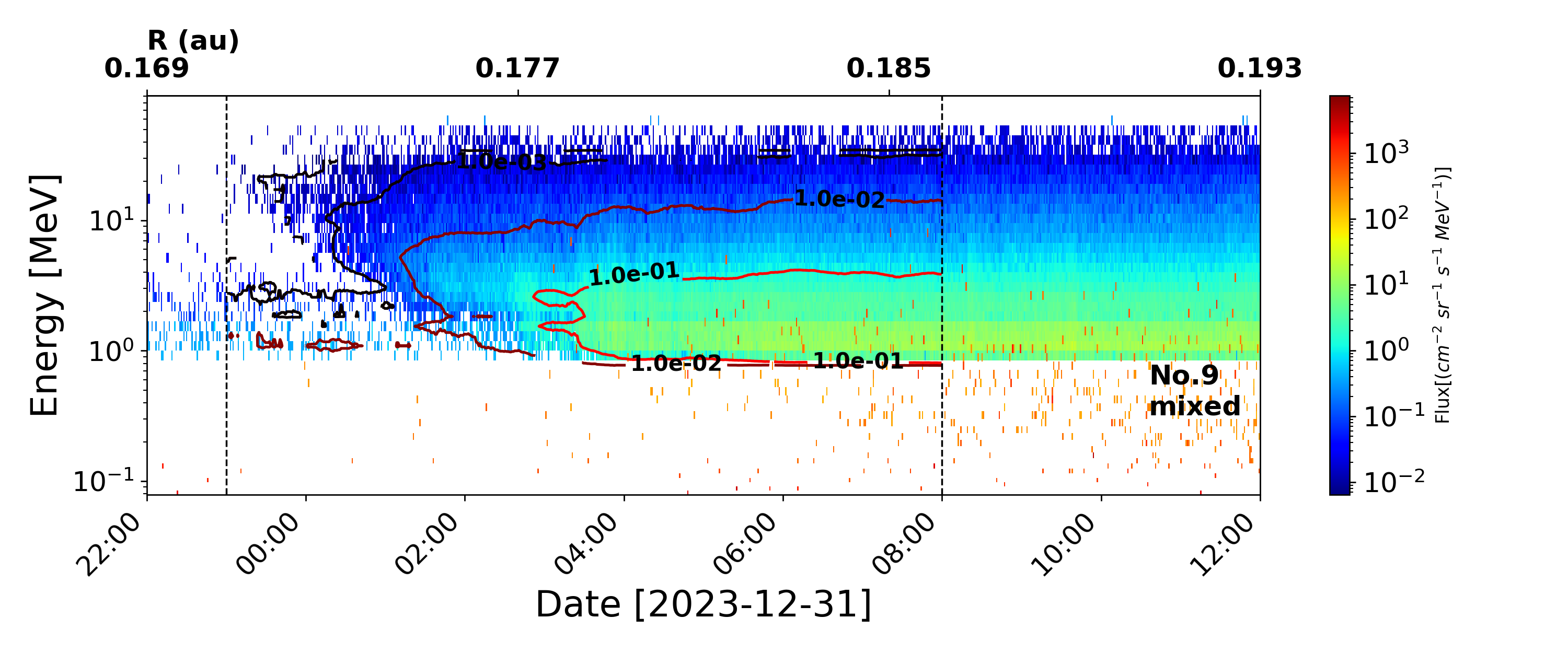}
    \includegraphics[width=0.4\textwidth]{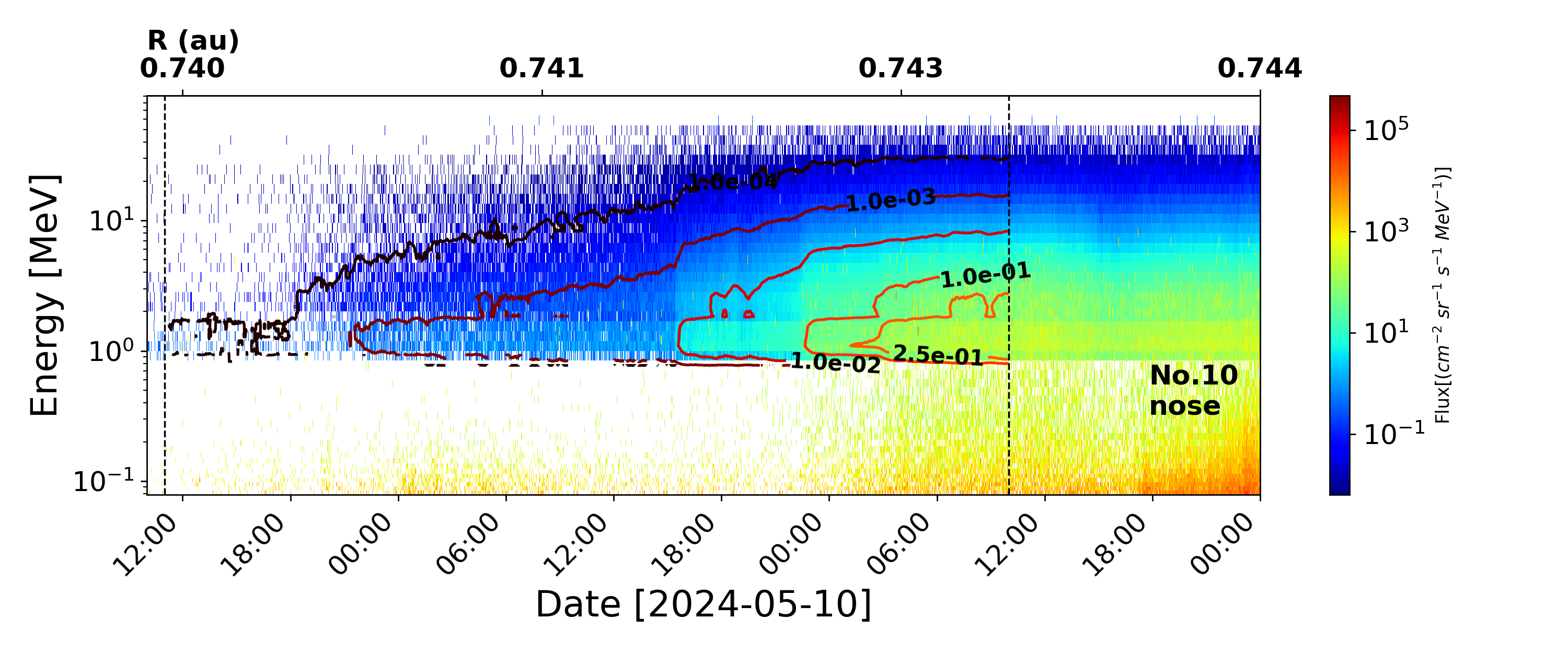}
    \includegraphics[width=0.4\textwidth]{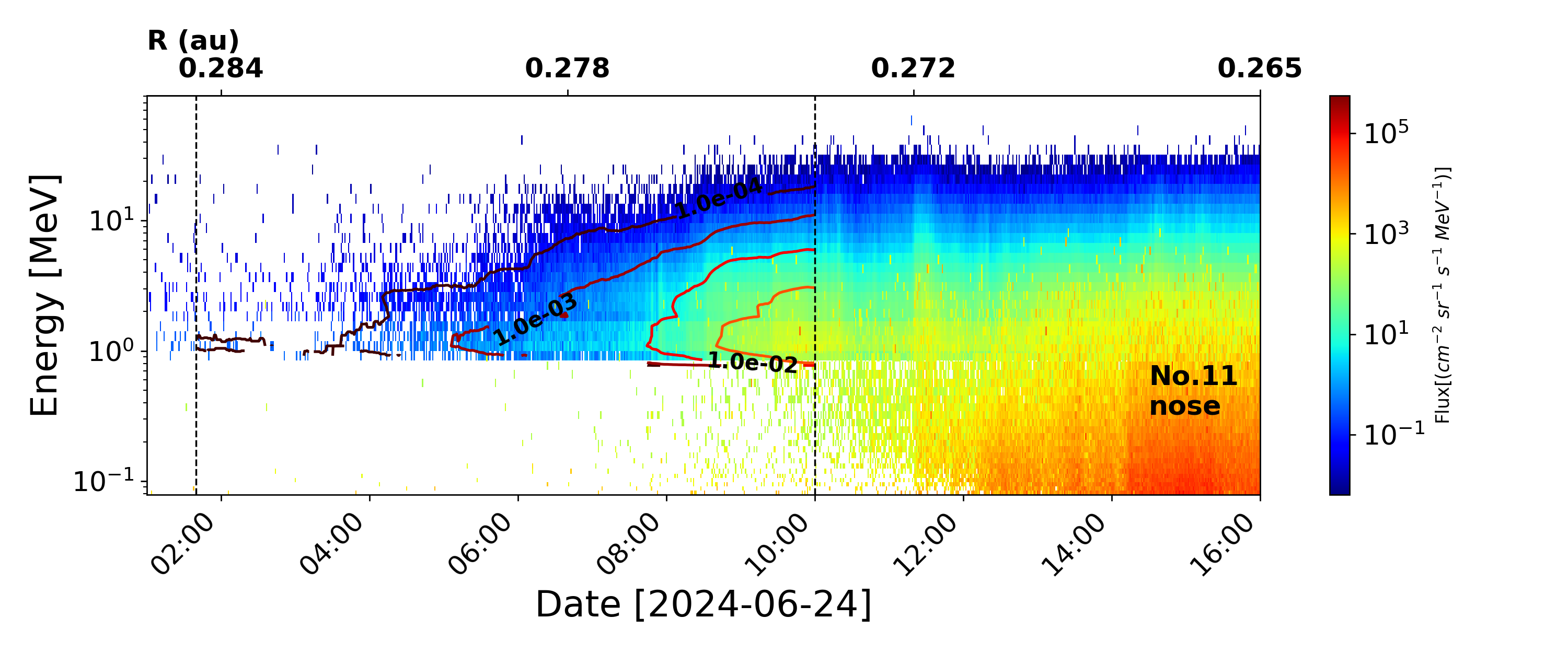}
    \includegraphics[width=0.4\textwidth]{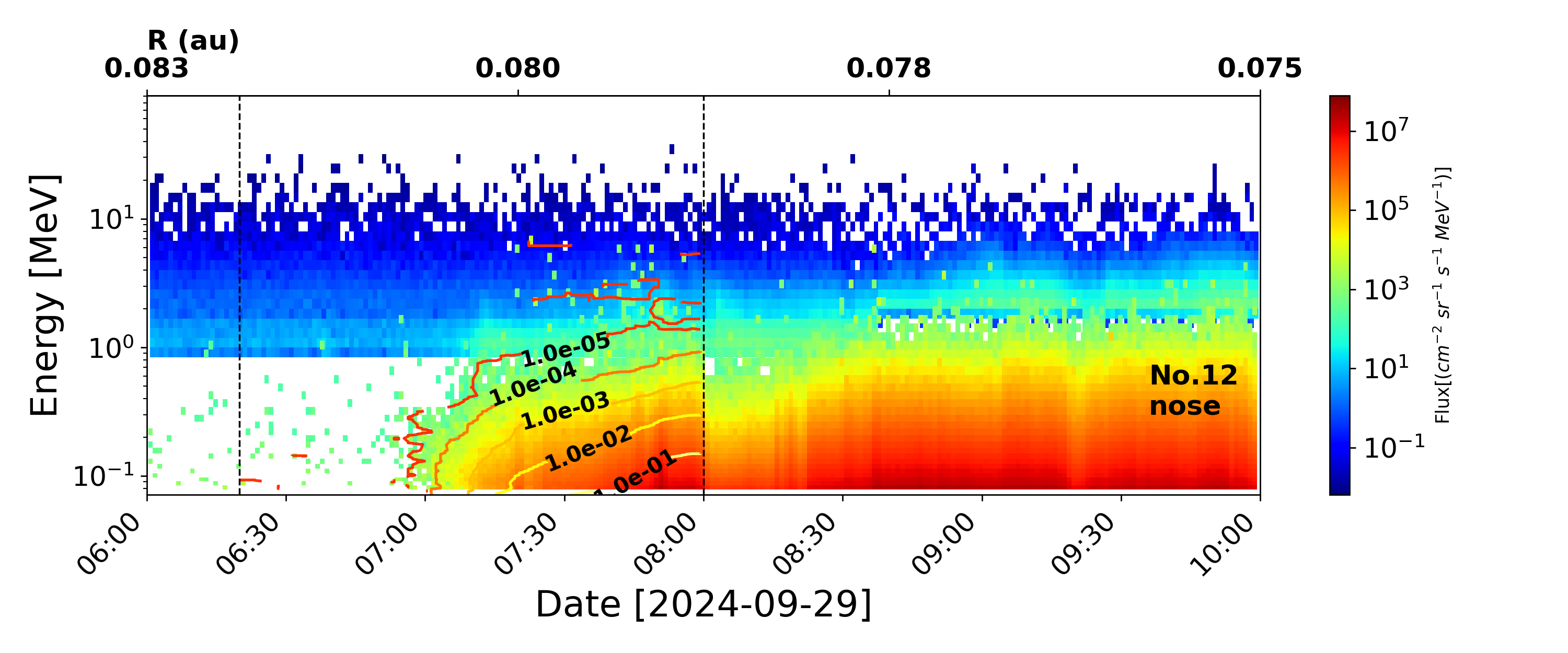}
    \includegraphics[width=0.4\textwidth]{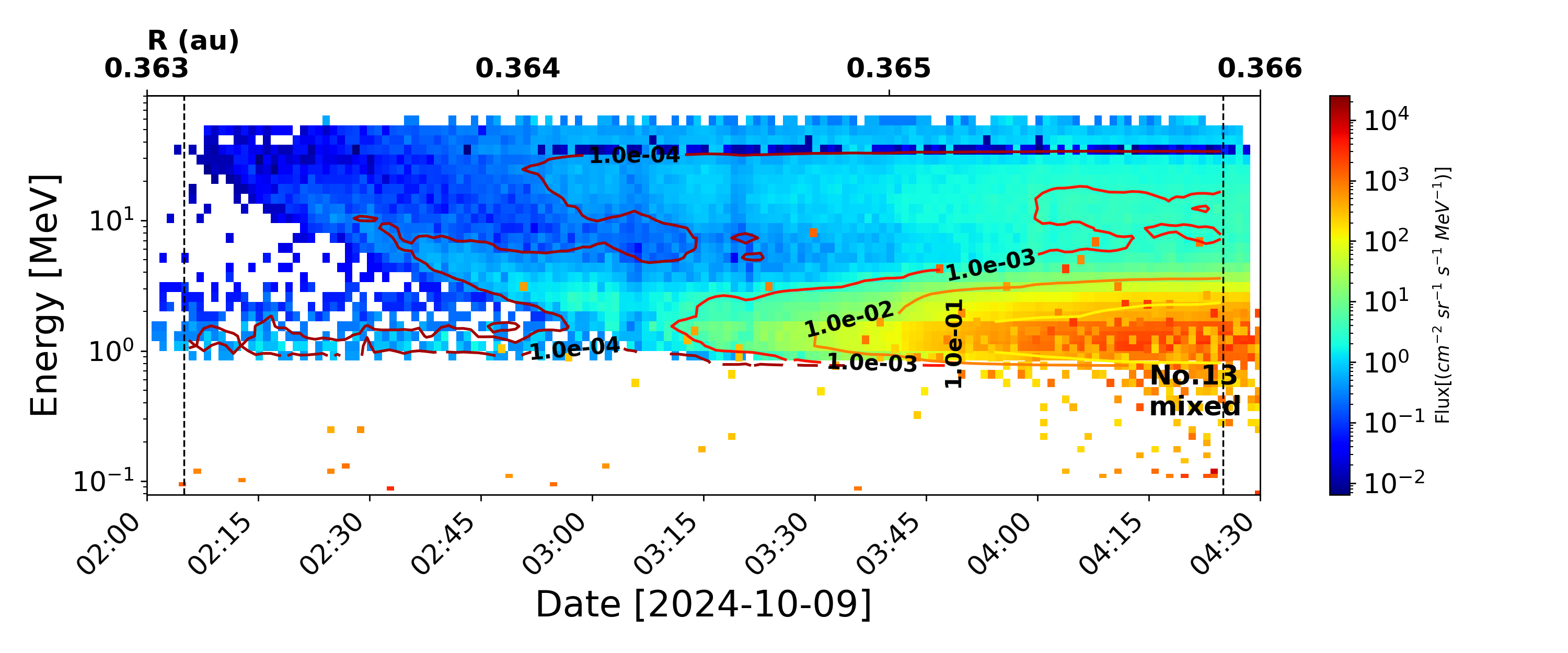}
    \includegraphics[width=0.4\textwidth]{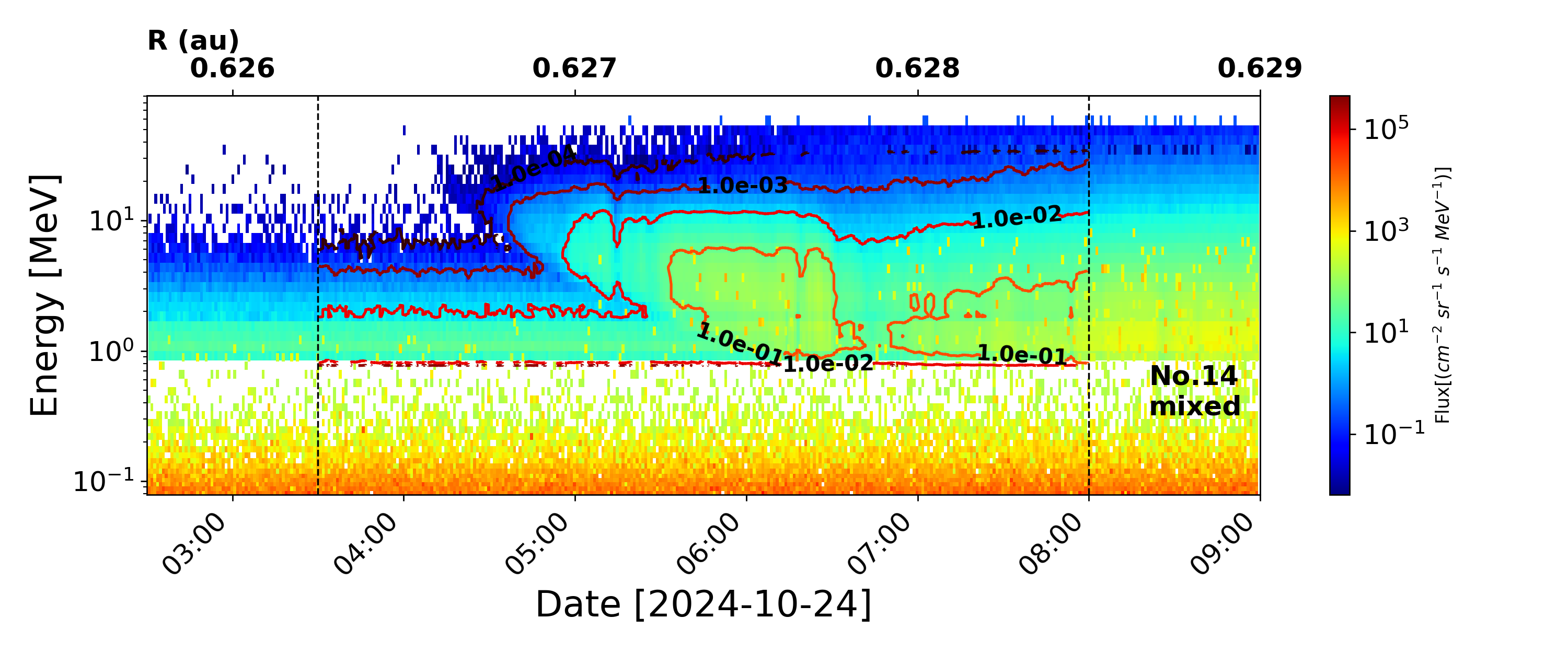}  
    \caption{The dynamic spectrograms of the other 12 IVA events. The contour lines are drawn between two vertical dashed lines.}
    \label{fig:SEP-2022-08-29}
    \label{fig:Appendix-IVA}
\end{figure}

\subsection{Properties}

\begin{figure}
    \centering
    \includegraphics[width=0.48\linewidth]{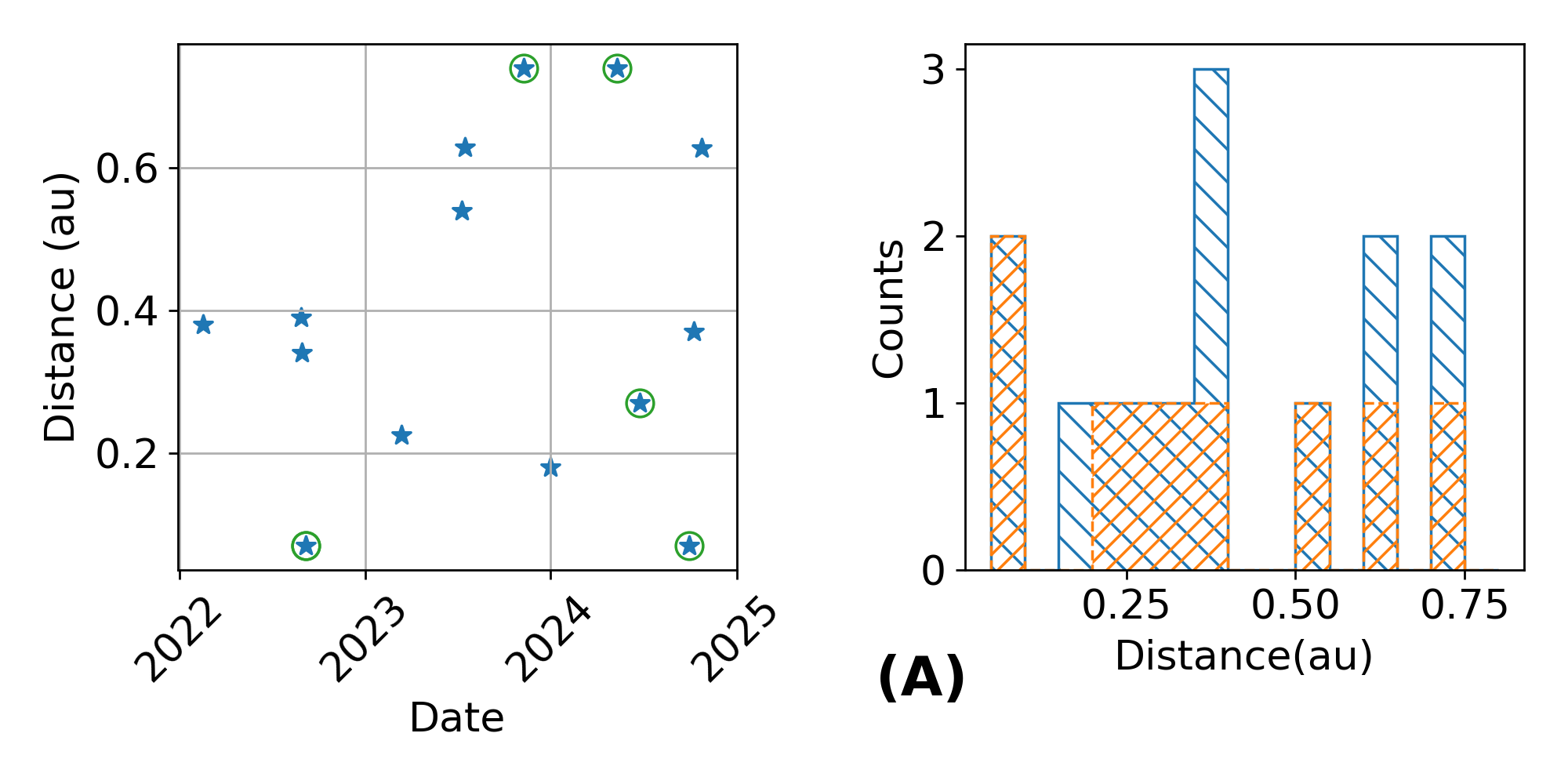}
    \includegraphics[width =0.48\linewidth]{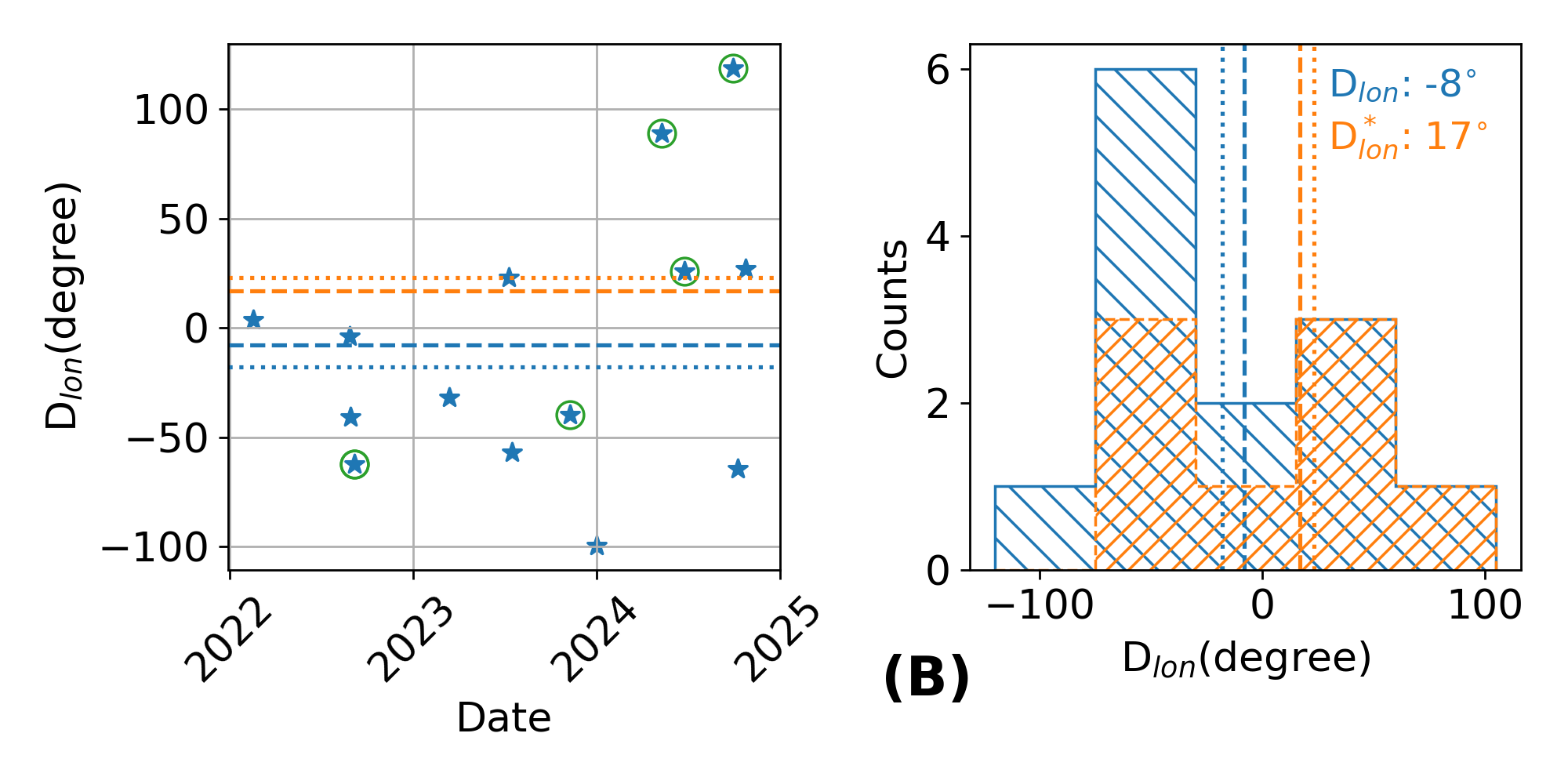}
    \includegraphics[width =0.48\linewidth]{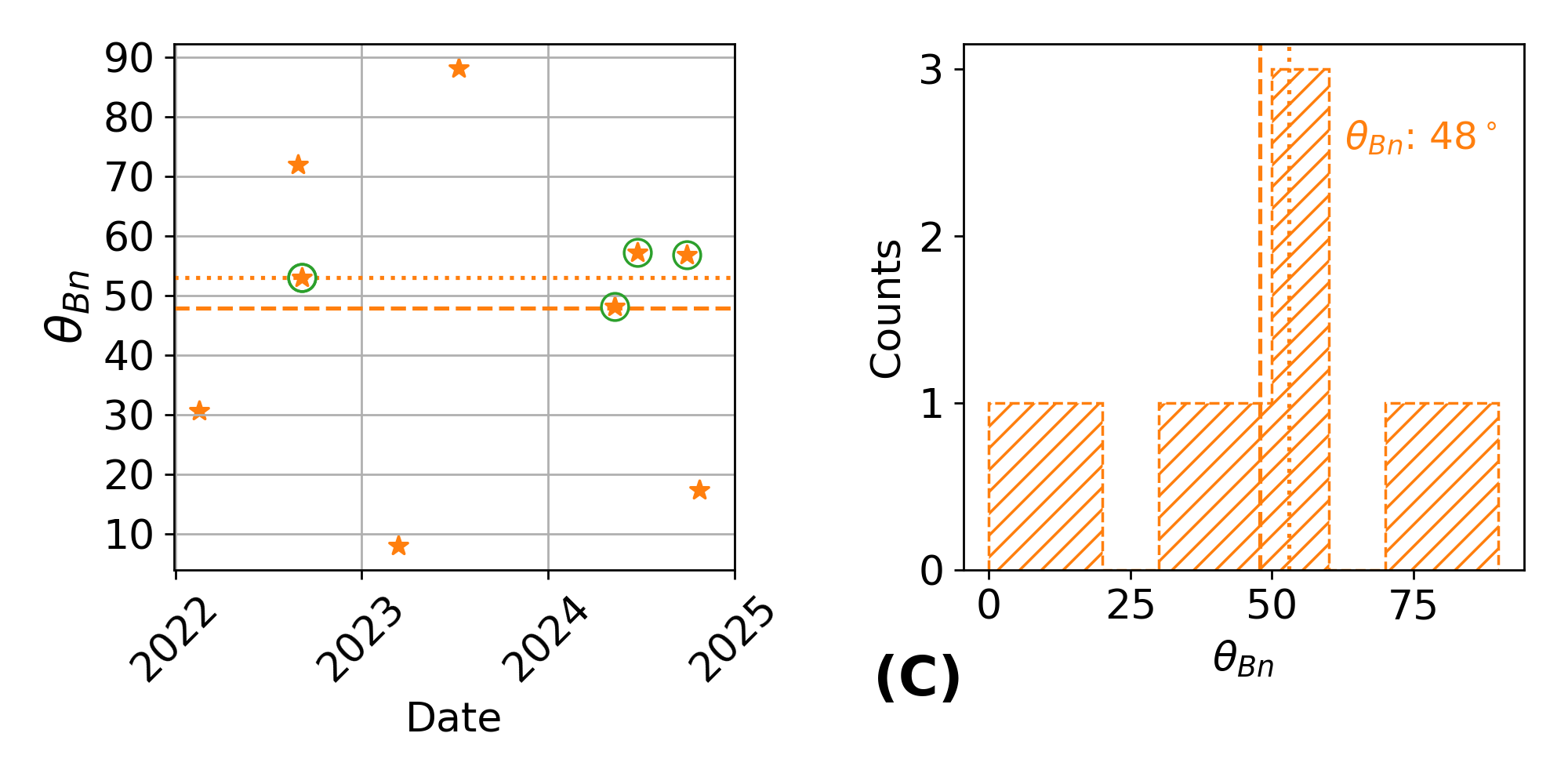}
    \includegraphics[width=0.48\linewidth]{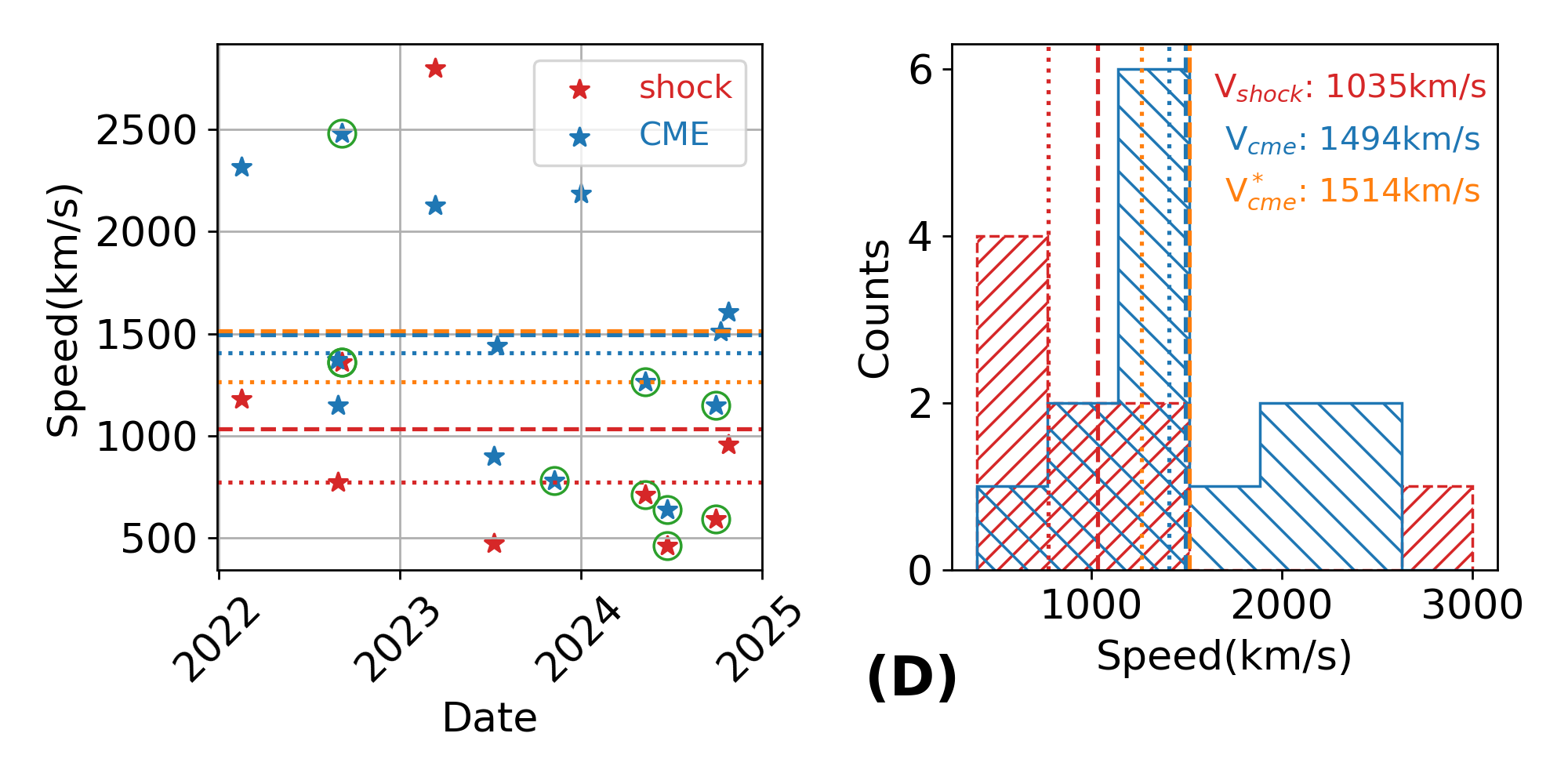}
    \caption{The distribution of the IVA events along the radial distance (A), versus the longitudinal separation from the flare (B), the $\theta_{Bn}$ of the shock (C), and the speed of the shock when it passes PSP and the speed of CME (D). The left sub-panels display the distribution in the order of time, and the right sub-panels present histograms of the distribution. The blue colored stars and histogram represent results for all 14 cases, while the orange colored ones are for events with shocks measured locally at PSP. The vertical dashed and the dotted lines indicate the corresponding mean and median values in the given panels. Specifically, red stars and histogram in panel (D) indicate the speed of local shock that arrived at PSP, and} the mean and median speeds are presented as red vertical lines. To maintain figure clarity, the histogram of CME speeds associated with local shocks is omitted, and only the averaged value is shown in the figure.
    The green circles in each panel indicate the nose-only events.
    \label{fig:shock_parameters}
\end{figure}
\begin{figure}
    \centering
    \includegraphics[width=\linewidth]{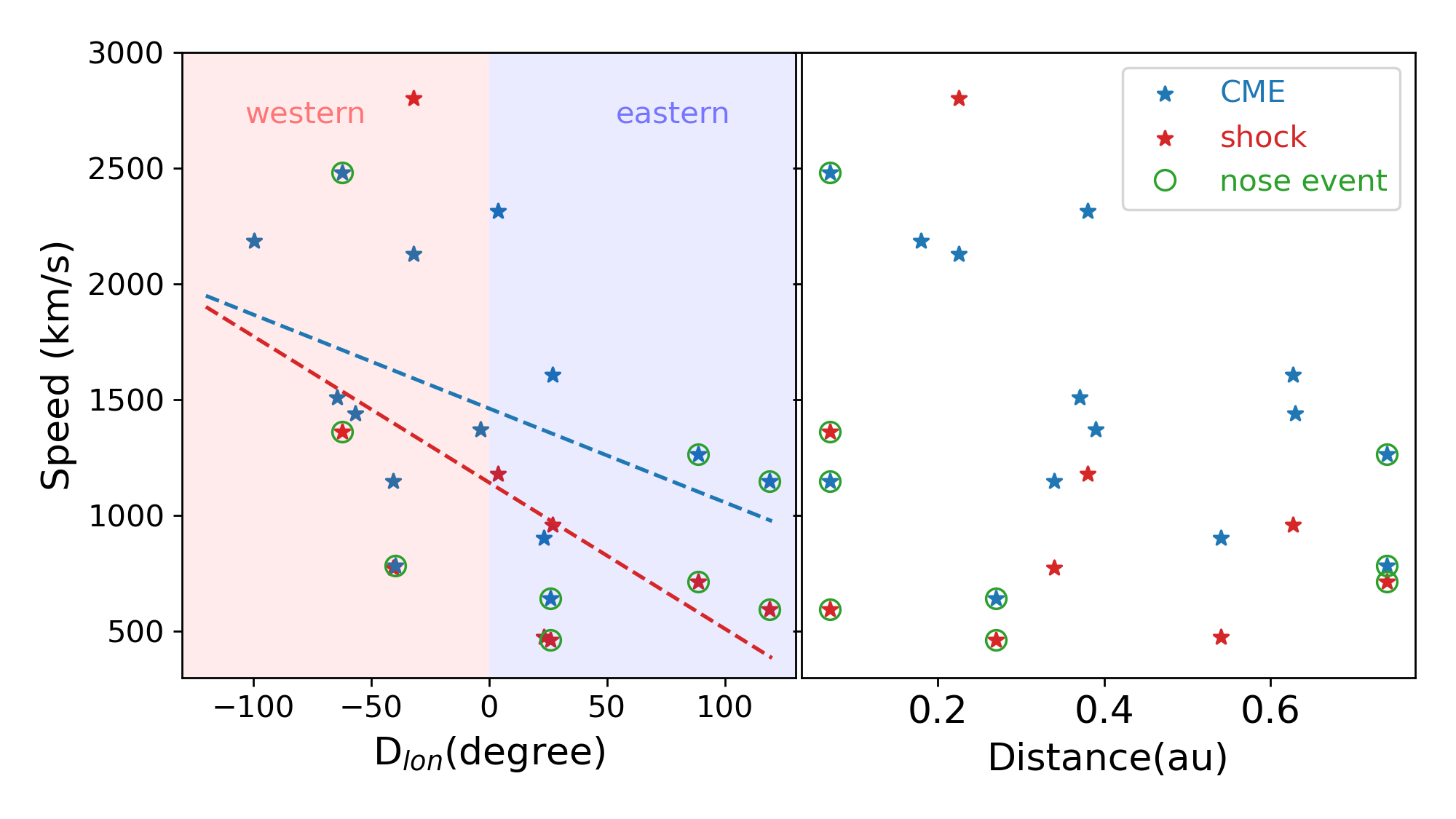}
    \caption{The relationship between the speed of the local shock (V$_{sh}$ ) / CME (V$_{cme}$) and the longitudinal separation from the spacecraft's magnetic footpoint to the flare, and the radial distance of the spacecraft, and }
    \label{fig:vsh-long-distance}
\end{figure}

We note that the following analysis of the characteristics of IVA events is affected by the limited number of events in our survey, hence the statistical generality of our results should be viewed with caution. The preliminary results are presented in Figures~\ref{fig:shock_parameters} and \ref{fig:vsh-long-distance}.

Figure~\ref{fig:shock_parameters} has four panels (A-D), and each panel has two sub-panels showing the results related to the radial distance (A), the longitudinal separation (B), $\theta_{Bn}$ (C), and the shock/CME speed (D).
The left sub-panels provide the values for the individual SEP events in the order of time, and the right sub-panels present histograms of the distribution. In all subpanels, the blue colored histograms are for all 14 cases, while the orange colored ones are for the 9 events with measurements of a shock at PSP.
The distribution of the IVA events as a function of radial distance in panel (A) indicates that IVA events can appear at any radial distance in PSP's range. The limited number of events cannot reveal an 'optimal' distance for such events, although there seems to be more cases closer to the Sun for the orange histogram. However, considering the limited time that PSP spends near perihelion, the weights of those IVA events near the Sun should be larger.

D$_{lon}$ indicates the longitudinal separation between the magnetic footpoint of PSP and the estimated flare location using the remote sensing data, which reflects the magnetic connectivity. Positive values indicate eastern SEP events in which PSP's footpoint is west of the flare. For events with large positive values, the spacecraft may be likely connected to the west flank of any CME-driven shock when the shock is close to the Sun. Events with negative D$_{lon}$ are western SEP events, with the magnetic footpoints of PSP east of the flare and intersecting CME-driven shocks on the eastern flank when such values are large. D$_{lon}$ values close to 0 degrees indicate a good magnetic connectivity to the center of the shock front (or the flare) when it is close to the Sun, which is typically expected to be an efficient acceleration region for quasi-parallel shocks, although dependent on the shock strength and geometry. It should be noted however, that the solar longitude where the PSP field line intersects the shock will be significantly different than the calculated spacecraft footpoint longitude when the shock is far from the solar surface. 
Our results indicate that for the SEP events with IVA in our sample, the mean separation is around -8 degrees. After excluding events without shock measurements, the value becomes positive at 17 degrees. Both values come with a large standard deviation ($\sigma$) of $\sim$ 60 degrees, indicating a wide distribution of the longitudinal separation, and thus are not statistically significantly different. The mean values are displayed as the dashed horizontal/vertical lines for both cases. The dotted lines represent the median values, which can minimize the impact of the outliers and here are consistent with the mean values.

In the second row, we present the shock parameters, including $\theta_{Bn}$ and the shock/CME speed. It is important to note that the shock parameters shown here correspond to shocks measured at PSP, rather than near the Sun. Obviously, in cases where PSP is located close to the Sun, the measured shock is expected to closely resemble the initial shock near the solar surface, though the shock parameters may vary along the shock surface. The CME speed is taken as an indication of the strength of the particle accelerator, i.e., the shock close to the Sun.
As shown in panel (C), we derive an average $\theta_{Bn}$ of 48 degrees, with most cases located between 40 and 60 degrees and $\sigma$ of 24 degrees. The median value is slightly above 50 degrees. This suggests that when the shock arrives at PSP, it is more likely to be an oblique shock.
In the last panel (D), we display both the shock speed when the shock arrives at PSP and the initial CME speed. The averaged shock speed (red dashed line) is around 1035 ($\sigma$=688) km/s, but with a lower median value, as the dotted red line indicates. The blue histogram bars show the distribution of the CME speeds.
These initial CME speeds are higher than the local shock speed, with a mean value of 1494 ($\sigma$=563) km/s (blue dashed line) for the whole dataset and 1514 ($\sigma$=617) km/s (orange dashed line) for the 9 cases with a local measurement of a shock. The median speeds are slightly less for all three values, indicating contributions from the smaller speeds. Typically, the shock speed should be significantly higher than the CME speed at a given location. However, the shock speeds we derived here are obtained when the shocks arrive at PSP, reflecting the local shock property, while the CME speeds are the speed of the CME when it is still close to the Sun. Therefore, it is understandable that our CME speeds are higher than the shock speed at PSP, which is decelerated due to the drag force from the local solar wind \citep{vrsnak_propagation_2013}. 

Obviously, no matter which parameters, the results are constrained by the limited number of samples, especially for the radial distribution; hence, future studies would highly benefit from more IVA events. 
We must consider some caveats on Figure~\ref{fig:shock_parameters}
and Figure~\ref{fig:vsh-long-distance} regarding CMEs and shocks. The CME speeds reported in the DONKI dataset represent the maximum values obtained from the model fitting and are the projected values in the plane of sky, depending on whether more than one coronagraph observed the events. Consequently, it is often unclear whether the derived speed corresponds to the CME apex or its flank. Therefore, the CME speeds have relatively large uncertainties. Similarly, the shock speeds derived from PSP can vary significantly with the spacecraft’s crossing location, whether at the flank or the center. Moreover, the derived shock parameters are not necessarily the same as when the initial particles were accelerated.
After propagating through the solar wind, a shock can substantially evolve, and its properties may largely differ from those it had when accelerating particles closer to the Sun.

Despite that, we examine how $\mathrm{V_{sh}}$ and $\mathrm{V_{cme}}$ correlate with the $\mathrm{D_{lon}}$ 
and spacecraft radial distance. The results are shown in the two panels of Figure~\ref{fig:vsh-long-distance}. While not conclusive, both the shock speed (red stars) and the CME speed (blue stars) seem to correlate with the longitudinal separation between the magnetic footpoint of PSP and the flare location with medium Pearson correlation coefficient values of -0.51 and -0.42, respectively. The two dashed lines are the simple linear regression, although the confidence interval remains limited due to the small number of events and unknown uncertainties of the CME speed and the shock speed. 
Again, because of our small sample size the tendency for western IVA events to have large CME and shock speeds should be regarded as suggestive and not statistically significant. A systematic comparison with the full population of SEP events is required to reliably quantify any longitudinal asymmetry, which is beyond the scope of this work. Regarding the radial distance, there seems to be no obvious correlation between distance and the speeds of the CMEs and shocks.



\section{Discussion}

The IVA signatures measured by PSP and Solar Orbiter suggest a previously unrecognized characteristic of SEPs observed within the inner heliosphere, particularly in the close perihelion regime that PSP traverses. These events occurred in a largely unexplored region with solar wind properties, plasma environment, magnetic field, and shock properties that have not been thoroughly investigated. 
Simultaneous presence of IVA at higher energies and VD at lower energies marks the onset (arrival) of the second population, i.e., the nose population. This population can occur in isolation, e.g., 'nose-only' events, or arrive after an earlier population exhibiting standard VD signatures to form 'mixed' events. Utilizing the contour line method, we have identified 14 IVA events in the PSP data through the end of 2024.

\subsection{The sensitivity of the instrument}
Before addressing the physical mechanism that may give rise to the IVA, we first discuss the instrumental effects on the observed spectrogram pattern of the IVA related SEP events arising from differential detection efficiency. As discussed by \citet{laitinen_correcting_2015}, the onset times at higher energies are later than those at lower energies when the peak intensities are less than 10 times the background levels, hence the resulted VDA are affected by the intensity ratio between the background and event.

We consider the Labor Day event as an illustrative case. As shown in Figure \ref{fig:2022-0905} and Figure \ref{fig:contourline-spectrogram-b}, the energy spectrogram of LET differs from that of  EPI-Lo, as manifested by the distinct inverted energy edge, where LET observed the inverted trend earlier than EPI-Lo. In addition, both the duration of the inverted energy trend and the slope of the edge differ between the two instruments.
The underlying reason is that EPI-Lo has a higher intensity threshold than  EPI-Hi/LET and HET, as reflected in their geometry factor values, which quantify the collection power of the instrument. 
As noted in Section~\ref{sec:instru}, the geometry factors of LET and HET are approximately an order of magnitude larger than the total geometry factor of EPI-Lo. Consequently, for the same accumulation period, LET and HET can measure particle intensities nearly an order of magnitude lower than those detectable by EPI-Lo.
Moreover, as SEP events progress, the EPI-Lo detection efficiency at the higher energies may further drop due to the increased instrumental deadtime caused by a huge amount of lower energy particles, hence worsening the discrepancy at different energy channels. 
The lower detection efficiency of EPI-Lo primarily affects the shape of the energy spectrogram and the apparent onset time at different energy channels, rather than the absolute intensity level.
As a result, the observed nose pattern at EPI-Lo represents only a portion of the complete profile. 
This can also be an issue between different particle instruments aboard different spacecraft, like PSP and Solar Orbiter.
One of our follow-up studies will investigate detailed cross-comparisons to further test our hypothesis that differences in detector efficiency can lead to variations in the observed nose pattern.





\subsection{Nose-only events vs mixed events}


Three types of SEP events, i.e., normal VD, nose only, and the mixed event, that we introduced in Sect.~\ref{sec:threetypical}, represent a new way of looking at SEP events. 
SEP events with normal VD correspond to the typical fast-rising SEP events due to good magnetic reconnection during flare eruptions or acceleration at the CME-driven shock when PSP is well connected to a region of strong acceleration efficiency, where most high-energy particles are accelerated and escaped into space via open magnetic field lines. 
In such events, the peak intensities are usually detected shortly after the onset of the SEP event, and the energetic particles forming the normal VD are the first observed population arriving at the spacecraft.
In the nose only SEP events and the mixed SEP events, the energetic particles that form the IVA features are a result of more gradual acceleration by CME-driven shocks to higher energies. 
The acceleration time required by higher energy particles, together with the connection time that the observer first magnetically connect to the shock region where particles are accelerated and escape since eruption, and the transport time of particles from the shock front to the observer, can account for the total delay in the arrival of this population at the spacecraft \citep{chen_evidence_2025}.

The mixed SEP events consist of both groups of particles, with the first population of particles having normal VD and the second population displaying both IVA and VD to form the characteristic. The overall appearance of the mixed event depends on the dominance of the two particle groups. When the first particle population significantly exceeds the second in intensity, the VD feature is more prominent than the nose feature, and the converse also holds true. The nose-only event and the VD-only event are thus the two extreme ends of a continuum. 
It is also possible that the first particle population is largely missed in our current observations due to its lower intensity and the limited sensitivity of the instruments.
In theory, a sufficiently sensitive telescope could measure a statistically significant number of lower-intensity particles over a short interval, and in that case, it is reasonable to expect that all SEP events would include first-arriving particles with a VD signature.

There are possibly hints of this in some of the nose-only events in our list. During the September 5, 2022 event, the onset of the SEP event was at 16:42 UT according to LET (see the second dashed line in Figure~\ref{fig:2022-0905}), whereas the onset of the event reported by \citet{cohen_observations_2024} at EPI-Lo is later than LET at 16:48 UT. Before that, the protons of even lower intensity had already arrived at PSP and were observed by LET, as exhibited by the scattered pixels in blue color. One may mistakenly consider these as background because of their limited statistics and lack of observations in EPI-Lo, which has a higher intensity threshold. However, compared to the pre-event background, the intensities at those few MeV energy channels are significantly enhanced. Also, by examining the LET energy spectrum from 1 MeV to 10 MeV, we found that the power-law index between the first two dashed lines in Figure~\ref{fig:2022-0905}, i.e., the eruption of the flare at 16:10 UT and the start of the event at 16:42 UT, is approximately -2.3, which is significantly steeper than the index of -1.3 measured during the quiet time one hour earlier. Thereafter, the spectrum further steepens to -3.8 at the start of the nose event between 16:42 UT and 17:10 UT, a period before  LET transitioned into DT mode due to the elevated intensities.

Additionally, the delay time between the flare eruption and the arrival of these earlier particles is reasonable. The radio observations in the bottom two panels suggest that the type-III radio burst starts around 16:10 UT, which is consistent with the peak time of the HXR emission at Solar Orbiter \citep{vievering_unraveling_2024} after considering the light travel time and the release time of the proton measured at Solar Orbiter \citep{kouloumvakos_shock_2025} as inferred from the velocity dispersion analysis.
The arrival time of these few MeV particles at PSP is roughly estimated as the halfway between the two dashed lines, i.e., about 16 minutes after the flare eruption. This time delay is very close to the minimum required time for a few MeV protons to travel across 0.07 au, which is less than 10 minutes. The slight difference between the two values is understandable considering the 60-degree longitudinal separation between the estimated magnetic footpoint of PSP and the flare region.  

This slight enhancement of the few MeV protons in the 2022 September 5 SEP event may be a small portion of a population of generally field-aligned protons that have moved across field lines to arrive at PSP.
Previous studies \citep{dresing_17_2023, Xu2020ApJ, khoo_multispacecraft_2024, dresing_reason_2025} on the cause of widespread SEP events have proposed several possible explanations, such as perpendicular diffusion, multiple injections, or the irregular magnetic field lines. Though by which means the particles travel through the extended space at the earlier phase of the 2022 September 5 SEP event is still unclear, the PSP observations may provide insights into how fast particles travel across field lines (unfortunately, this is out of the scope of this paper). Alternatively, these earlier-arriving protons are accelerated at the flank of the shock, a region that is presumably weaker than the central part of the shock. Their acceleration rate is limited by the seed particle population and the shock geometry, resulting in proton intensities that are lower, and with lower energies, than the later-arriving protons.

The 2022 September 5 nose-only event is a so far unique SEP event in which the IVA feature is observed by both EPI-Hi and EPI-Lo with a sharp intensity onset. We have not found any similar events with such a strong and clear indication of the inverse feature.
The closest contender in our current survey is the 2024 September 29 event \citep{wilson_large-amplitude_2025}, whose spectrogram is given in Figure ~\ref{fig:Appendix-IVA}. 
The most distinct characteristic of these two events compared to other SEP events is their close distance to the Sun and to the particle acceleration site. Both of them occurred during a close perihelion at around 0.07 au (15.45 vs 17.07 R$_s$).
Although both of them have nose features, their nose energies are very different. The nose energy of the Labor Day event is around a few MeV, while the SEP event on 2024 September 29 has a very low nose energy below 0.5 MeV, and is only visible by EPI-Lo. Further, the accompanying CME-driven shock of the 2024 September 29 SEP event is not as strong as that of the Labor Day event.



It has long been argued that SEP events are caused by different mechanisms, reconnection or acceleration, and from distinct sources, i.e., flare or CME-driven shock \citep{krucker1999origin, reames_two_2013, desai_large_2016}. Many studies have discussed their roles in shaping SEP events from different perspectives. For instance, \cite{Xu_2024ApJ} presents a SEP event case in which observers at different longitudinal locations saw particles from different sources, as reflected in the observed compositional differences. However, during any large SEP event, the contributions from different sources may still be unclear, since they often both occur and can accelerate particles to higher energy.
Therefore, examination of compositional differences between two populations may be revealing, though we have to admit that it might be challenge to do this due to limited statistics of heavy ions.








\subsection{Event context}
According to the prediction of the time-dependent acceleration scenario, the shape of the nose can be explained by the balance between the particle acceleration time and its travel time, particularly when the former is not negligible relative to the latter. The energy of the first-arriving particles is related to the acceleration efficiency of the shock, which is a crucial parameter in many astrophysical contexts. At the present stage, determining the location of the nose energy remains subject to large uncertainties, in particular for the mixed events. Moreover, the nose energy is event-dependent, since each SEP event occurs under different physical conditions. Despite that, statistical results indicate that most IVA related SEP events have characteristic nose energies between 0.5 MeV and 5 MeV.

Apart from the Labor Day event, all other medium-energy IVA SEP events were observed when PSP was at a distance beyond 0.3 au, according to our survey results. A larger distance between the observer and the shock results in a longer particle travel time and thus, in order to maintain a balance between the acceleration and transport time scales, the efficiency of the shock acceleration for higher energy particles must be substantially reduced compared to similar events observed when the shock is much closer to the spacecraft.  As this may be difficult to achieve, an alternative explanation related to the observer's evolving magnetic connection to the shock and the shock's non-uniformity in acceleration efficiency should be considered.  As described in \cite{ding_investigation_2025}, when the magnetic connection between the spacecraft and the shock progresses from a region of low acceleration efficiency to one of higher efficiency a nose feature may be created.  This effect is a less plausible explanation for near-Sun nose events, such as the Labor day event, as the evolution in connection would likely be minimal for shocks very close to the spacecraft. Investigating the relative importance of these two scenarios requires study and careful modeling of the shock and spacecraft observations in many more events. 

Another intriguing feature is the separation between the PSP's magnetic footpoint and the flare location. Though several cases indicate a large longitudinal separation on both west and east sides, the average values of the full IVA event list and those with local shocks only are -8 and 17 degrees, respectively. Both distributions exhibit relatively large standard deviation of $\sim$ 60 degrees, indicating a broad spread in longitudinal separation.
Given the current small and heterogeneous sample, it is difficult to draw a firm conclusion about the preferred locations of the associated flare and CMEs, though we believe the magnetic connectivity should play a key role in the generation of the IVA feature. The observed distributions suggest that an IVA event may appear at either side of the flare longitude, with a possible preference toward the eastern side of the flare, as the blue histogram in panel (B) indicates. This trend appears to contradict the finding from Solar Orbiter \citep{allen_delayed_2026}, where they reported a clear tendency for IVA events observed by Solar Orbiter to occur when the spacecraft magnetic footpoint was on the western side of the associated flare location. The reason behind this contradiction is still being investigated, but the hypothesis could be that Solar Orbiter IVA events occur due to non-uniform shock acceleration efficiency across the shock front \citep{ding_investigation_2025}.

IVA is manifested as a change in the acceleration efficiency of the shock from the beginning phase to the later phase of acceleration 
in which it is capable of accelerating high-energy particles after an extended acceleration time \citep{li_energetic_2003}. Therefore, the acceleration regions to which the observer is connected cannot be the very high compression regions that could more easily accelerate particles to very high energies in a short time. In other words, the IVA signatures are indicative of a less efficient acceleration process early in the event but one that increases as the event evolves, compared to the most energetic SEP events.

However, we need to keep in mind that the later high-energy particles could also be due to continued acceleration in which  the acceleration region in a non-uniform shock front evolves from a less efficient acceleration region to a better one \citep{ding_investigation_2025}.
Such a change might be in any shock property, e.g., compression ratio, Mach number, and the shock geometry, all of which can affect the acceleration rate over time and change as the CME/Shock expands \citep{kouloumvakos_effect_2023}.

\subsection{Is IVA an ESP event near the sun?}

Within the overall framework of IVA event generation due to the time-dependent shock acceleration, the IVA feature may be detectable when the observer is in close proximity to any given shock. In this sense, although this has not been examined systematically, IVA events may share similarities with ESP events, which arise from particles accelerated locally at or shortly before shock arrival.
Observations of ESP events have shown an inverted edge in the energy spectrum with the approach of the CME/shock \citep{lario_evolution_2019,lario_high-energy_2023}. Though the exact mechanisms that produces IVA-like signatures are still not well established, the similarity of inverted spectrograms may imply a common process associated with the shock en route to the observer.
In addition, for some IVA events, the particle acceleration may occur at distances quite close to the spacecraft. For instance, during the Labor Day event, the shock reached PSP about half an hour after the energetic particle onset. The IVA pattern of the 2023 March 13 event was only two hours before the shock arrival. 

But the differences are also obvious. During an ESP event, the normal VD at lower energies is generally not present, and the profiles are manifested as an overall intensity enhancement across an extended energy range. 
The other distinction between them is the relative location of the observers with respect to the particle acceleration region. Traditional ESP events are considered as the result of local acceleration and particle are observed almost simultaneously with minimal transport effects, while the partilce propagation time associated with IVA events is non-negligible. In addition, the ages of the CMEs in  ESP and IVA events, and possibly the seed populations are also likely different. How those factors impact the two kinds of events and whether or not the ESP events reflect the same mechanism as IVA events is worth further investigation.

\section{Summary}

We introduce a more systematic framework to identify and analyze the IVA feature using the contour line method. By examining past observations from PSP between 2018 and the end of 2024, we report a total of 14 IVA related SEP events.
We propose a two-population scenario to understand the SEP events. The first population is the normal VD particles, and the second population has both VD and IVA, creating a nose feature. The IVA pattern is associated with the CME-driven shock due to its time-dependent acceleration nature and is dependent on the shock properties. In any give event the relative strength of those two populations can yield a VD only, nose only, or mixed type of event. By identifying the two populations, we may be able to provide a new methodology for separating the contribution of the different accelerators, and determine which part or which factor is most dominant for each population. 

Using this framework we have categorized SEP events observed by PSP into three types: VD, nose-only, and mixed events. Here we examined events that had a clear nose population (either as a nose-only or mixed event type) and identified 14 events.
For these 14 events, we found that the averaged longitudinal separation (D$_{lon}$) between the magnetic footpoint of PSP and the flare location is between -8$^\circ$ and 17$^\circ$, but with very broad distributions (e.g., standard deviations of $\sim$ 60 deg). The mean $\theta_{Bn}$ and velocity of the CME-driven shocks when they pass the spacecraft are around 48$^\circ$ and 1035 km/s, respectively. The shock and the CME speed exhibit a weak correlation with the D$_{lon}$, however without more study, it is not clear if this is a feature particular to IVA events. No preferred occurrence distance in the heliosphere can be identified based on our current observations. Regarding the nose energy, which is the energy of the first arriving particles of the nose population, we found that most of the IVA events have a medium nose energy between 0.5 and 5 MeV.


Looked at as an ESP event, the IVA events are associated with shock acceleration occur at various locations. 
IVA events have been observed and reported by multiple missions and observers, including PSP, Solar Orbiter, and BepiColumbo. The better energy resolution and coverage of the new particle telescopes, and the novel way of visualizing the particle profiles using dynamic spectrograms, enabled the discovery of this inverse feature. In the future, a promising research direction will be to explore the enormous historical dataset from multiple instruments at 1 au, like STEREO, to determine whether similar phenomena are present. Regardless of whether similar events are identified, such an investigation would provide valuable insights into the characteristics of IVA events and their relationship to local shock acceleration and ESP events. Instrumental discrepancies, of course, must also be carefully considered.





\begin{acknowledgments}

We acknowledge the contribution of the Solar Orbiter and Parker Solar  Probe mission teams, especially those of the Energetic Particle Detector and the \ISOIS \ instrument teams when analyzing the data. Parker Solar Probe was designed, built, and is now operated
by the Johns Hopkins Applied Physics Laboratory (JHU/APL)
as part of NASA’s Living with a Star (LWS) program (contract
NNN06AA01C). Support from the LWS management and
technical team has played a critical role in the success of the
Parker Solar Probe mission. We thank the scientists and
engineers whose technical contributions prelaunch have made
the \ISOIS \ instruments such a success. A.K. acknowledges financial support from NASA NNN06AA01C (PSP EPI-Lo) contract and NASA’s LWS grant 80NSSC25K0130. RCA acknowledges support from NASA contract 80GSFC25CA035 and NASA grant 80NSSC24K0908.
\end{acknowledgments}

\bibliography{main}{}
\bibliographystyle{aasjournal}



\end{document}